\documentclass[pra,aps,twocolumn,showpacs,amsmath,amssymb,superscriptaddress]{revtex4}
\usepackage{amssymb,amsmath}
\usepackage{color}
\usepackage{graphicx}

\begin{document}

\title{Spatial fluctuations in an optical parametric oscillator below
threshold with an intracavity photonic crystal}

\author{M.~A. Garcia-March}
\affiliation{Department of Physics, Colorado School of Mines, Golden, CO, 80401 }
\affiliation{Department of Physics, University College Cork, Cork, Ireland}
\affiliation{Institute for Cross-Disciplinary Physics and Complex Systems, IFISC
(CSIC-UIB), Palma de Mallorca, 07122, Spain}
\author{M. M. De Castro}
\affiliation{Institute for Cross-Disciplinary Physics and Complex Systems, IFISC
(CSIC-UIB), Palma de Mallorca, 07122, Spain}
\author{R. Zambrini}
\affiliation{Institute for Cross-Disciplinary Physics and Complex Systems, IFISC
(CSIC-UIB), Palma de Mallorca, 07122, Spain}


\begin{abstract}
We show how to control spatial quantum correlations in a
multimode degenerate optical parametric
oscillator type I below threshold by introducing a spatially inhomogeneous medium, such
as a photonic crystal, in the plane perpendicular to light propagation. We obtain the
analytical expressions for all the correlations in terms of the relevant parameters of
the problem and study the number of photons, entanglement,
squeezing, and twin beams. Considering different regimes and configurations
we show the possibility to tune the
instability thresholds as well as the quantumness of correlations by breaking the
translational invariance of the system through a photonic crystal modulation.
\end{abstract}

\maketitle

\section{Introduction}

Quantum correlations between components of a system separated spatially or temporally
are on the foundations of whole new technological fields like quantum information and
computation, quantum cryptography or quantum-enhanced
metrology~\cite{2002GalindoRMP,polzik,2002GisinRMP,metrology}.  Many successful
implementations have been developed in optical systems in continuous variables
\cite{polzik,2005BraunsteinRMP}.  An important
example is the intensity of light far from the single photon regime.  The optical
parametric oscillator (OPO), a device in which a  classical input (laser) beam  is
down-converted  in a nonlinear medium, operates in many-photon regime. 
In these devices, the nonlinearity of the quadratic crystal allows for light modes
interactions, and, therefore, it is at the origin of the generation of
squeezing~\cite{1986WuPRL,2008VahlbruchPRL} and
entanglement~\cite{1992OuPRL,2005VillarPRL,2009CoelhoScience} in fields quadratures
\cite{2005BraunsteinRMP,polzik} (whose spectrum is continuous).
Different light modes can be distinguishable, for
instance, for their polarization or frequency.  Recently, special attention has been
devoted also to spatial degrees of freedom  where quantum correlations are displayed
between  cavity modes or parts of light beams~\cite{1999KolobovRMP}.  Indeed, many
applications have been already realized with multimode light, such as optical
switching~\cite{2005DawesScience}, quantum
imaging~\cite{2008BoyerScience,2009MarinoNature,2009JanousekNature,2010BridaNatPhotonics},
metrology~\cite{2003TrepsScience}, and quantum
information~\cite{2005OsullivanPRL,2007LassenPRL}.

A very appealing possibility is to use spatial inhomogeneities in broad area devices
to control relevant quantum properties of the light. In related fields, like linear or nonlinear
classical optics, it is well-known that the periodic  modulation of the refractive index
leads to gaps in the allowed frequencies of the electromagnetic field, a phenomenon
known as photonic band-gap~\cite{2008Joannopoulos,1987YablonovitchPRL,1987JohnPRL}.  As
a result, these engineered media, photonic crystals (PC), allow to confine and guide
light leading to many
applications~\cite{2008Joannopoulos,2007BuschPR,2003RussellScience}.  If a transverse
modulation of the refractive index is considered in nonlinear cavities, this was
predicted  to inhibit modulation instabilities at similar
wavelengths~\cite{2004GomilaPRL,2005GomilaPRE}. Recently this prediction was confirmed
experimentally~\cite{2008MarsalOL,2008TerhalleAPL} and a related phenomenon
was proposed in semiconductor microcavities~\cite{polaritons}.  Modulation in
dissipative systems leads also to the formation of nonlinear structures, such as
different kinds of discrete cavity solitons~\cite{2004PeschelOL}. Furthermore, the use
of PC is also at the basis of many proposals in quantum optics. Seminal papers
\cite{1987YablonovitchPRL} pointed at the possibility to inhibit spontaneous emission in
the PC band-gaps and this was recently observed in different experiments 
\cite{Noda2007NatPhot,2004LodahlNature,2005FujitaScience}. As a matter of fact, the use
of PC for environment (dissipation) engineering, stemming from the presence of photonic
band-gaps, is the basis of intense research activity on cavity QED. In particular,
non-Markovian effects have been predicted in quantum optics with  structured
reservoirs~\cite{2000LambropoulosRPP}, exploring also effects on decoherence dynamics
and entanglement decay~\cite{2008PiiloPRL,2007BellomoPRL}.

In this work we show the effect of a transversal modulation on quantum fluctuations and
correlations in a nonlinear device where the presence of gaps is expected to inhibit
quantum fluctuations. We consider a photonic crystal optical parametric oscillator
(PCOPO),  that is, a multimode degenerate OPO with a PC in the cavity, as described
in~\cite{2005GomilaPRE} where the PC effect on the modulation instability was studied
and compared with the homogeneous case~\cite{1994OppoPRA}.  In Ref.~\cite{2011DeCastro} we
presented a first study based on the numerical analysis  of quantum fluctuations
 using a Langevin treatment valid both above and below 
threshold~\cite{2003zambriniEPJD}. We  showed that the quantum correlations can be tuned by means
of this PC, obtaining noise reduction in field quadratures, robustness of squeezing in a
wider angular range, and, most remarkably, an improvement of entanglement above
threshold~\cite{2011DeCastro}.  In this paper, we present analytical results valid below
the parametric threshold and based on linear and few-modes  approximations in good
agreement with numerical simulations of the full model. We calculate intensity fluctuations
and correlations as well as quadratures squeezing and entanglement showing the effect of
the modulation. The homogeneous multimode OPO was shown to present squeezing,
entanglement and twin beams correlations between spatial modes below 
threshold~\cite{1993Lugiato,1997GattiPRA,2003ZambriniPRA} and  above threshold in presence of
stable patterns  or even frozen chaos~\cite{2003zambriniEPJD,2006PerezArjonaEPL}.
Similar effects have been predicted in Kerr media~\cite{2000ZambriniPRA} and in second
harmonic generation~\cite{2002BachePRA} and in the last years there have been several
successful experimental realizations
\cite{2008BoyerPRL,2008BoyerScience,2009MarinoNature,2009JanousekNature,2010BridaNatPhotonics,2010ChalopinPRA,2009LassenPRL,2011ChalopinOE}.
The effects of a spatial modulation here discussed for an OPO can also be generalized to
these other nonlinear devices modified by the inclusion of an intracavity PC.

The paper is organized  as follows. In Sec.~\ref{Sec:model}, we present the model of a
PCOPO using  linear and few modes approximations for the light fluctuations below
threshold giving the output signal field in terms of the input one. We also introduce a
set of non-linear Langevin equations~\cite{2003zambriniEPJD,2011DeCastro} that we
numerically simulate to test our approximations. In Sec.~\ref{Sec:threshold}, we obtain
an analytical expression for the intensities of the signal field, showing how the
instability threshold for signal emission can be either raised or lowered by  means of the PC spatial
modulation. Then, in Sec.~\ref{Sec:Qcorr} we obtain the expression for different quantum
correlations,  such as squeezing, entanglement and twin beams correlations. Last
section is devoted to our conclusions.

\section{Few mode approximation for the PCOPO}
\label{Sec:model}

We consider a planar cavity filled with a non linear $\chi^{(2)}$ medium
with one of the mirrors only partially reflecting. The pump field at frequency
$2\omega$ is down-converted in a signal at frequency $\omega$, with polarization
orthogonal to the pump one. This constitutes an implementation of a type I degenerate
OPO.  The input beam is a plane wave propagating in the
z-direction (the cavity axis)  with amplitude $E$, assumed real.  Here, we consider the
effect of the transversal inhomogeneity of the medium filling the cavity. This
corresponds, for example, to the introduction of a planar photonic crystal (PC) with
refractive index modulation in the plane perpendicular to the light propagation
direction. A sketch of the device is provided in Fig.~1 of Ref.~\cite{2011DeCastro}.
The intracavity dynamics of this PCOPO can be described in terms of a continuous of boson spatial modes
$\hat{A}_{0,1}(\mathbf{x},t)$ at frequencies $\omega_{0,1}$,
$\mathbf{x}\in\mathbb{R}^{2}$. These operators obey equal time commutation 
relations~\cite{1999KolobovRMP}:
\begin{equation}
\left[\hat{A}_{i}(\mathbf{x},t),\hat{A}_{j}^{\dagger}(\mathbf{x}',t)\right]=
\delta_{ij}\delta(\mathbf{x}'-\mathbf{x}).\label{eq:conmutation}\end{equation}
The Hamiltonian operator reads~\cite{1997GattiPRA,2003zambriniEPJD}
\begin{eqnarray}
\label{Eq:Hamiltonian}
\hat{H}=\hbar\gamma \int d^{2}\mathbf{x}\sum_{i=0,1}
\left[\hat{A}_{i}^{\dagger}\left(\Delta_{i}\left(\mathbf{x}\right)-
c_{i}\nabla^{2}\right)\hat{A}_{i}\right]+\nonumber\\  \label{eq:H}
iE\left(\hat{A}_{0}^{\dagger}-\hat{A}_{0}\right)+i
\frac{g}{2}\left(\hat{A}_{0}\hat{A}_{1}^{\dagger2}-
\mathrm{h.c.}\right)
\end{eqnarray}
where the first term describes diffraction of the fields in the cavity, with $\nabla^{2}$ the
Laplacian in the transverse plane and diffraction strengths $c_1=2c_0$. The second term accounts
for the interaction with the external pump $E$ while the nonlinear interaction between both
modes is given by the third term, being the coupling constant $g$   proportional to the
second-order susceptibility $\chi^{(2)}$. The coefficient $\gamma$ is the cavity damping rate
(introduced for convenience as a scaling).

The main difference with respect to a generic OPO~\cite{1997GattiPRA,2003zambriniEPJD}
is that in a PCOPO the intracavity photonic crystal gives rise to a spatial modulation
of the cavity detunings $\Delta_{0,1}(\mathbf{x})$~\cite{2011DeCastro}. This constitutes
a breaking of the translational symmetry of the system with deep consequences
both in the macroscopic fields dynamics and in the correlations between fluctuations. In
Ref.~\cite{2011DeCastro} numerical results about quantum effects both below and above
threshold where reported, based on simulation of the quantum fields
dynamics in the $Q$-representation. From the methodological point of view this
description, first discussed in Ref.~\cite{2003zambriniEPJD},  allows one to take into
account the full nonlinear dynamics with the drawback of being not amenable for
analytical calculations. In the following we introduce a
simplified and approximated model that we use in order to obtain analytical results
below the instability threshold.

The intracavity fields operators $\hat{A}_{0},~\hat{A}_{1}$
obey the Heisenberg equation
\begin{equation}
\label{Eq:HeisenbergEq}
\frac{\partial\hat{A}_{j}}{\partial t}=\frac{i}{\hbar}\left[\hat{H},\hat{A}_{j}\right]-
\gamma\hat{A}_{j}+\sqrt{2\gamma}\hat{A}_{j}^{\mathrm{in}}.
\end{equation}
where the dissipative contribution characterizes such an open system,
with $ \sqrt{2\gamma}\hat{A}_{j}^{\mathrm{in}}$ incoming quantum fluctuations
\cite{1991Gardiner}.   Due to the cubic form of the Hamiltonian (\ref{eq:H}),
the dynamic equations for the operator moments form an infinite hierarchy of coupled equations, 
which in turn is unsuitable to handle analytically.

A commonly invoked approximation based on a system size expansion
is the  linearization around a macroscopic steady state
leading to a dynamical evolution for the quantum fluctuations  governed by a {\it quadratic}
Hamiltonian~\cite{1991Carmichael}. 
Let us identify in each field operator a reference average value ${A}_{j}$
and a small fluctuation around it, $\hat{a}_{j}=\hat{A}_{j}-{A}_{j} $.  The reference values
 ${A}_{j}$ are the expectation values of $\hat{A}_{j}$ and their evolution
is obtained by averaging the Heisenberg equations
(\ref{Eq:HeisenbergEq}) and by approximating all nonlinear terms as the product of first
order moments. This procedure leads to  two classical equations whose steady state
clearly depends  on the regime in which the PCOPO is considered. If we consider pump
values $E$ such that the PCOPO is below the instability threshold, then the signal
operator expectation value is vanishing $A_1=0$, independently on the presence
of the PC. The equation for the average value of the pump field in this
regime reduces to
\begin{eqnarray*}
\partial_t A_{0}& = & -\left(1+i\Delta_{0}(\mathbf{x})-i\nabla^{2}\right)A_{0} +
E,
\end{eqnarray*}
where we have introduced the scaling $\mathbf{x'}=\mathbf{x}/\sqrt{c_1}$  and $t'= \gamma t$. 
This scaling is used in the remaining of this article together with the scaling  for the 
field variables described in~\cite{2003zambriniEPJD}. In the following we omit the primes to 
simplify notation.

In the case of homogeneous detuning $\Delta_{0}(x)=\Delta_{0}$ the steady state
solution of this equation
is immediately found. Then for an OPO (or a PCOPO whose modulation is only in the signal
detuning) the steady state is homogeneous, $A_{0}=E/(1+i\Delta_0)$.

For a non-homogeneous pump detuning the identification of the stationary state is
generally  not trivial. In the following, for the sake of simplicity, we consider only
one transverse  dimension and a PCOPO sinusoidal modulation  such that
\begin{eqnarray}\label{delta0}
\Delta_{0}(x)=\Delta_{0}+M_{0}\sin(k_{p}x)
\end{eqnarray}
where $k_{p}$ is the PC
wave-number. The steady state then satisfies, in the Fourier space,
\begin{eqnarray*}
\left(1+i\Delta_{0}+ik^{2}\right) {A}_{0}^{s}(k)&+&\frac{M_{0}}{2}\left( {A}_{0}^{s}(k-k_{p})\right. \\
 -\left. {A}_{0}^{s}(k+k_{p})\right) =  \delta(k)E.
\end{eqnarray*}
This gives rise to coupled mode equations for varying $k$. We neglect terms with
$|k|>k_{p}$, a key assumption justified in Sec.~\ref{sec:fewmodes}. We then obtain for the pump field
three non vanishing modes
\begin{equation}
{A}_{0}^{s}(x)=\sum_{k=0,\pm k_p} {A}_{0}^{s}(k)e^{ikx}
\end{equation}
with
\begin{eqnarray}
\label{eq:stationaryk}
 {A}_{0}^{s}(k_{p})&=&- {A}_{0}^{s}(-k_{p})=\frac{\frac{-M_{0}}{2} {A}_{0}^{s}(0)}
 {1+ik_{p}^{2}},
\\\label{eq:stationaryk0}
 {A}_{0}^{s}(k=0)&=&\frac{E\left(1+ik_{p}^{2}\right)}{1+ik_{p}^{2}+M_{0}^{2}/2}.
\end{eqnarray}
where we assume $\Delta_{0}=0$ without loss of generality. The steady states of the pump
and (vanishing) signal fields ${A}_{0,1}^{s}$ are  then a reference state about which
the fluctuations operators $\hat{a}_{0,1}$ are defined. With a standard procedure the
exact Hamiltonian is  approximated to one quadratic in these 
fluctuations~\cite{2003ZambriniPRA,1991Carmichael}. Pump and
signal Heisenberg equations are actually decoupled and the following dynamical equation
for the signal fluctuations is obtained:
\begin{eqnarray}
\partial_t \hat{a}_{1} & = & -\left(1+i\Delta_{1}-i2\nabla^{2}\right)\hat{a}_{1}+
 A_0^s \hat{a}_{1}^{\dagger}+\sqrt{2/\gamma}\hat{a}_{1}^{\mathrm{in}},\nonumber
\end{eqnarray}
\begin{figure}
\includegraphics[width=8cm]{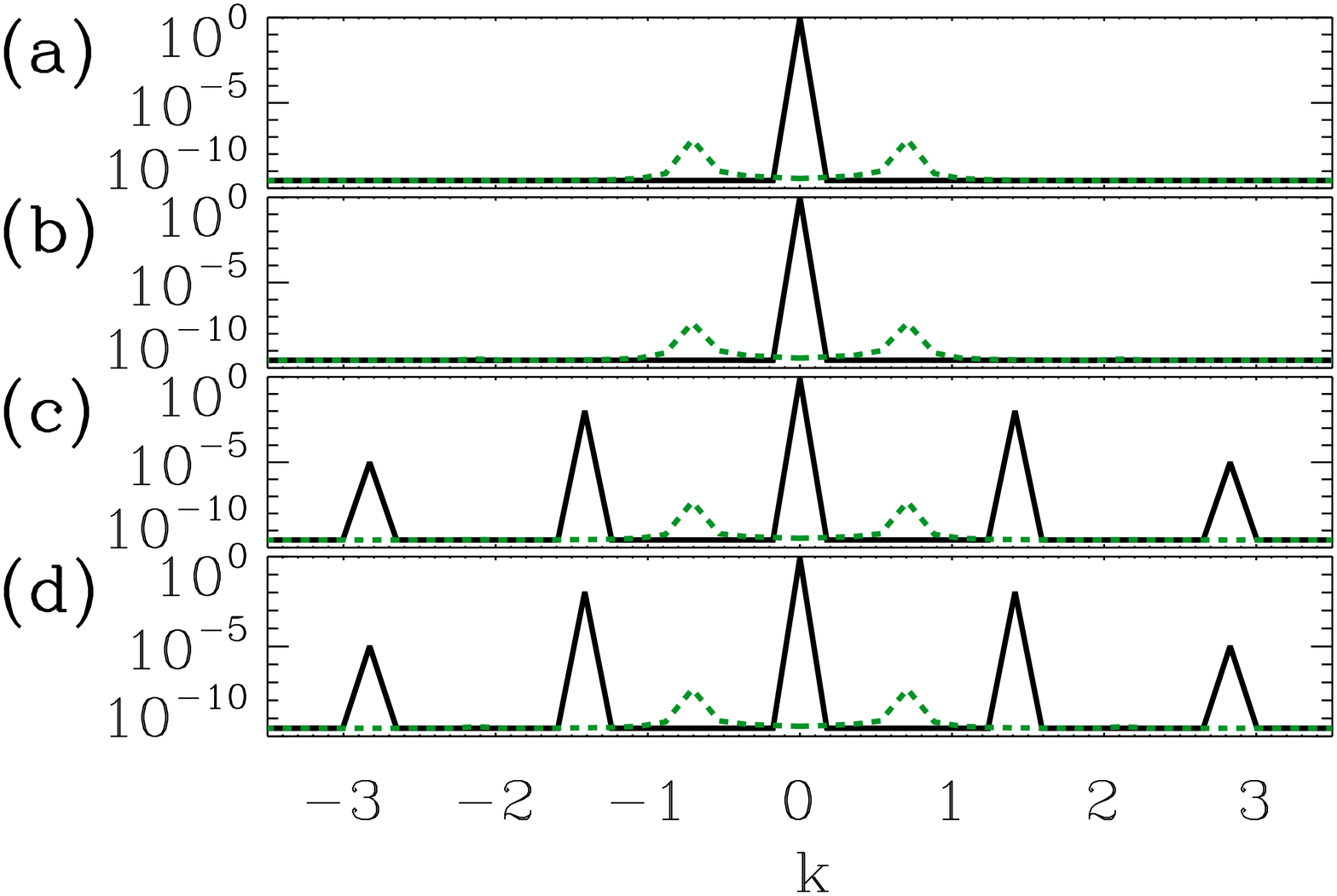}
\caption{(Color online) {\it Average far fields obtained by numerical simulation of the full
nonlinear Langevin equations}.
Pump in black solid line and signal in green dashed line. (a) corresponds to the case
without PC and  $E=0.999$, similar to the case with the PC affecting only the signal, 
represented in (b) for $M_1=0.5$, $M_0=0$, and $E=1.028$.   (c) shows the case with $M_0=0.5$, $M_1=0$,  and $E=0.931$,
similar to the case where the PC affects both fields, represented in  
(d) for $M_0=M_1=0.5$ and $E=0.956$.   The homogeneous component is
always present and harmonics appear for the signal at $k_c=\sqrt{-\Delta_1/2}\simeq 0.7$ and
for the pump at $2k_c$. The external pump is a $0.1\%$ smaller than its corresponding threshold. 
\label{fig1}}
\end{figure}

To simplify notation in the following we drop the hats of the operators
and we denote the fluctuation $\hat{a}_1$ as $a_1$.
The modulation on the signal will be similar to the pump one
\begin{eqnarray}\label{delta1}
\Delta_{1}(x)=\Delta_{1}+M_{1}\sin(k_{p}x).
\end{eqnarray}
Notice that, in general, the amplitudes of both modulation are not equal, $M_{1}\neq M_0$.
Due to the presence of the PC, the far fields fluctuations operators
${a}_{1}(k)$
\footnote{For
the Fourier transform, we use the convention
$a(k,\omega)=\frac{1}{2\pi}\int d^{2}x\frac{1}{\sqrt{2\pi}}\int dt e^{-i(kx-\omega
t)}a(x,t)$ and, for the hermitian conjugate, $a^{\dagger}(-k,-\omega)=\frac{1}{2\pi}\int
d^{2}x\frac{1}{\sqrt{2\pi}}\int dt e^{-i(kx-\omega t)}a^{\dagger}(x,t)$.}
 do not evolve independently. Different (k wave-vectors or) modes are dynamically
 coupled
\begin{eqnarray}
\label{Eq:enFourier}
& &\partial_t{a}_{1}(k,t)  =  -\left(1+i\Delta_{1}+i2k^{2}\right)a_{1}(k,t)\nonumber\\
& &+\frac{M_{1}}{2}\left(a_{1}(k+k_{p},t)-a_{1}(k-k_{p},t)\right)\\
&+&\sum_{n=0,\pm1 }{A}_{0}^{s}(nk_{p},t)a_{1}^{\dagger}(-k+nk_{p},t)+
\sqrt{\frac{2}{\gamma}}a_{1}^{\mathrm{in}}(k,t),\nonumber
\end{eqnarray}
due to the spatial modulation of both the signal detuning
($M_1\neq 0$) and the pump one
[through the spatial harmonics $ {A}_{0}^{s}(nk_{p}) $]. Notice that
neglecting
higher harmonics is equivalent to consider only $n=0,\pm 1$ in Eq.~(\ref{Eq:enFourier}).
Still, Eq.~(\ref{Eq:enFourier}) unveils the
 dynamical  coupling of $6$ different modes, as discussed in
App.~\ref{Sec:BigL}, and
further approximations are needed in order to handle this
model analytically.

\subsection{Numerical simulations of fully multimode and nonlinear dynamics}

This leads us to examine the full PCOPO model in order to identify the most relevant
spatial modes in different regimes. 
The full dynamics can be calculated  by numerical
simulation of Langevin equations~\cite{2003zambriniEPJD,2011DeCastro},  obtained by
mapping the full master equation for the PCOPO --whose system Hamiltonian is given in
Eq.~(\ref{Eq:Hamiltonian})--  onto an equation of motion for the Husimi quasi-probability
distribution $Q$ in phase space~\cite{1991Gardiner,2003zambriniEPJD}.  This
representation is then a functional  of the $c$-number fields $\alpha_i({x}) $ that are used to get
the expectation values of the operators $\hat{A}_i( {x})$\cite{1991Gardiner}.
In regimes where pump intensities are not too high,
the Husimi distribution Q dynamics is governed by a Fokker-Planck equation,  mapped in the following
nonlinear Langevin equations for spatial dependent pump $\alpha_0$ and signal $\alpha_1$
fields~\cite{2003zambriniEPJD}:
\begin{eqnarray}
\label{Eq:lang}
\partial_t
\alpha_0( x ,t)&=& - \left[(1+i\Delta_0( x ))-i\nabla^2 \right]\alpha_0(
x,t)+ \nonumber  \\ \nonumber && E-\frac{1}{2}\alpha_1^2( x ,t)+\xi_0( x ,t)\\ \nonumber
\partial_t
\alpha_1( x ,t)&=& - \left[(1+i\Delta_1( x ))-2i\nabla^2
\right]\alpha_1( x ,t)+\\ && \alpha_0( x ,t)
\alpha_1^*( x ,t)+\xi_1( x ,t).
\end{eqnarray}
with $\xi_0$ additive and $\xi_1$ multiplicative, {\it phase sensitive}, white noises.
Notice that
again the effect of the  PC is enclosed in the spatial dependence of the detunings $\Delta_0(x)$ and
$\Delta_1(x)$.

To study the dynamics of the system  we have simulated these equations numerically (technical details
about numerical methods are given in Ref.~
\cite{2002BachePRA}).  For the OPO, without  PC, it is known that, for negative signal
detuning, a modulation instability develops above the parametric threshold ~\cite{1994OppoPRA}, with
critical wavenumber $k_c=\sqrt{-\Delta_1/2}$ (here we fix $\Delta_1=-1$ so that $k_c\simeq 0.7$).
Below threshold, the pump is homogeneous while noisy precursors at the critical wave-number are
observed, with an average intensity increasing when approaching the threshold
~\cite{2003zambriniEPJD}. In Fig.~\ref{fig1}a we show the average pump and signal far fields in the
OPO, with the homogeneous and critical modes excited, respectively.

We now consider the PCOPO and in particular we focus on the case in which the  PC has a
band-gap $at$ the critical wave-number
of the OPO
\begin{eqnarray}\label{2kc}
k_{p}=2k_{c}
\end{eqnarray}
If only the signal is modulated, for $M_0=0$ and $M_1=0.5$, the PCOPO shows average far fields
similar to the case of the OPO. On the other hand,
if we introduce a modulation in the pump detuning, $M_0\ne0$, the pump field
develops many even harmonics
$ 2k_p,4k_p,6k_p...$, as shown in Fig.~\ref{fig1}c and d. It is also shown that
the signal average far field intensity
remains unchanged in all cases.

\begin{figure}
\begin{tabular}{cc}
\includegraphics[width=4cm]{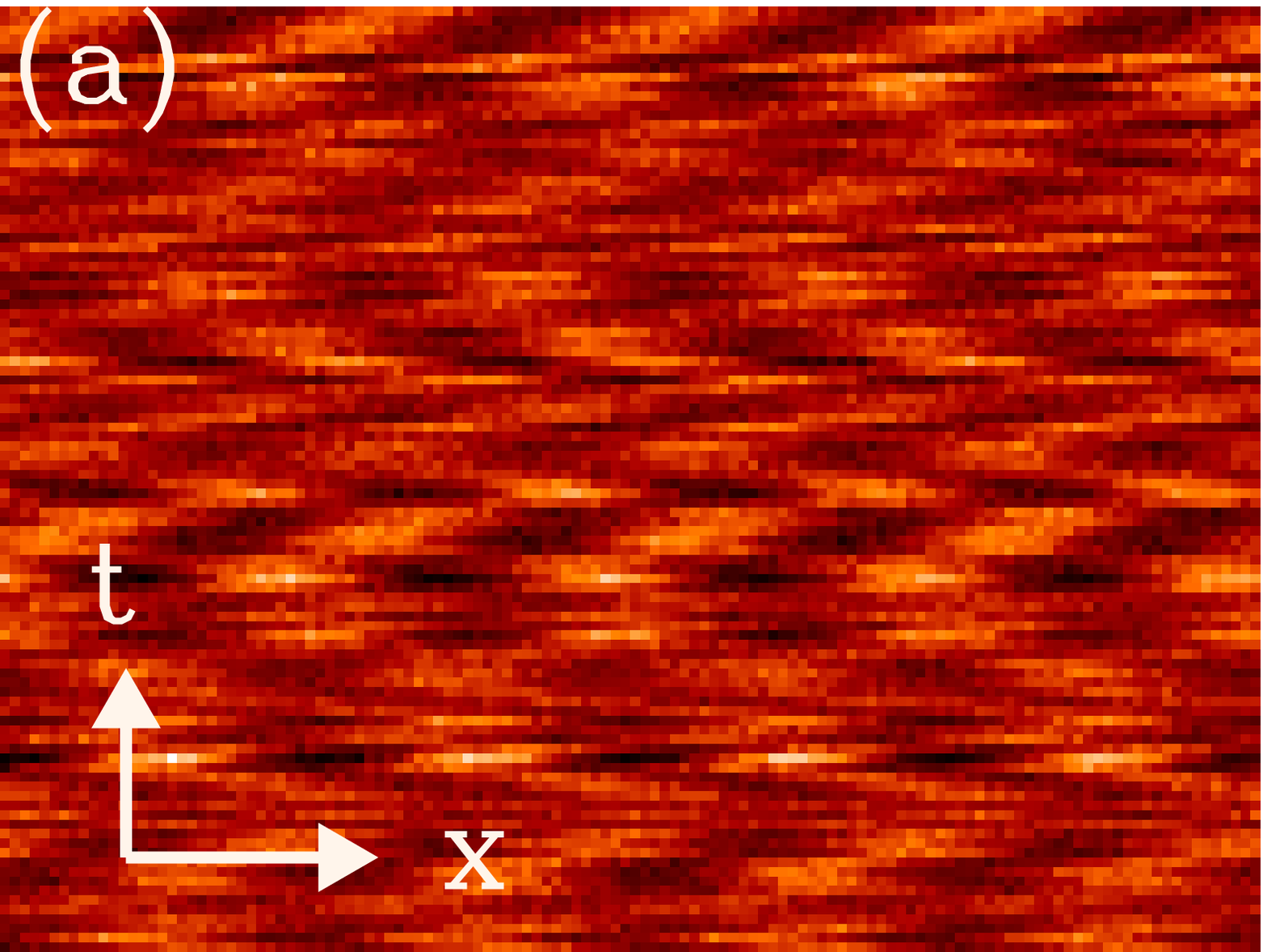}&
\includegraphics[width=4cm]{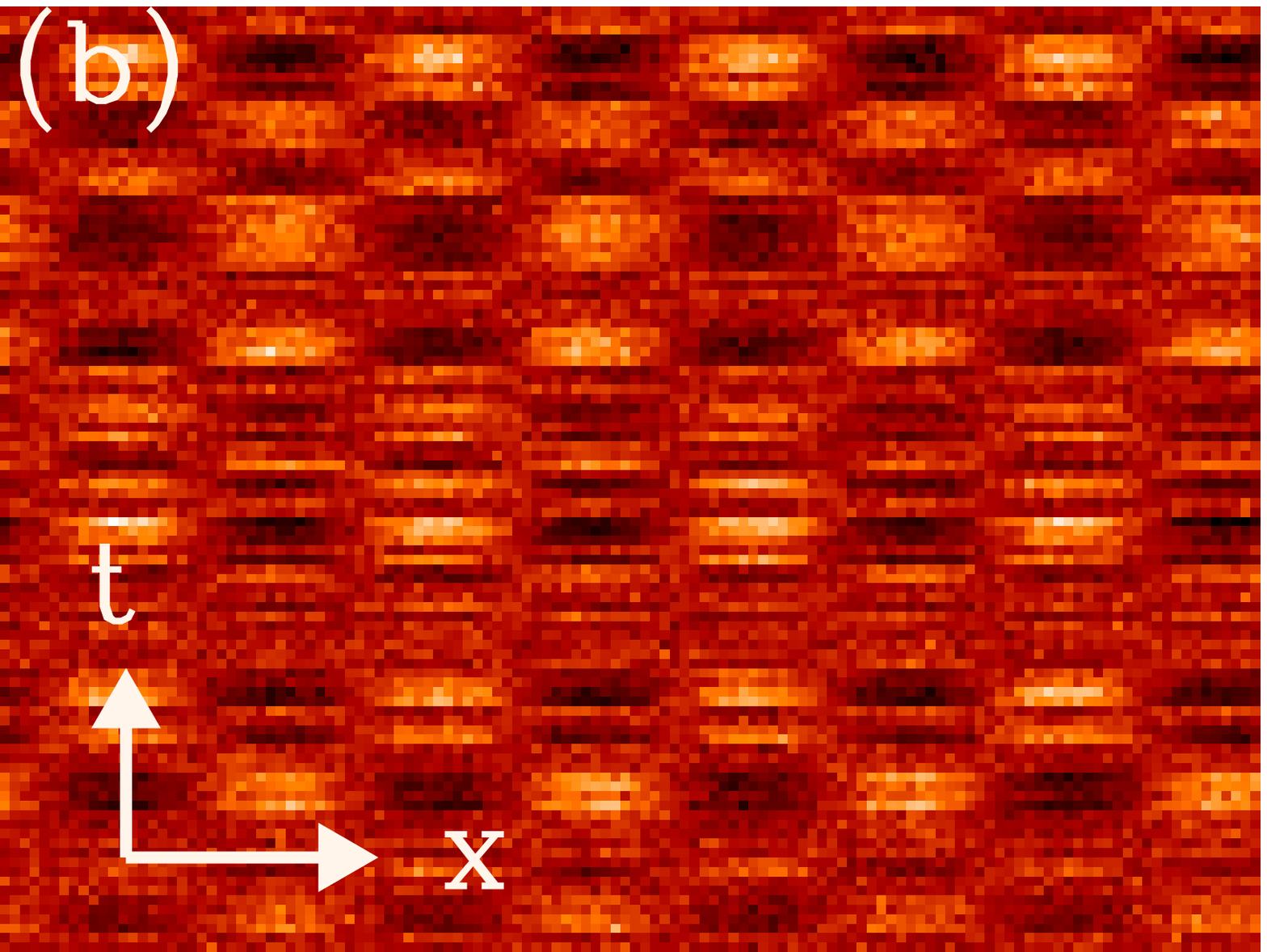}
\end{tabular}
\caption{(Color on line)  Near field evolution of the real part of signal $\alpha_1(x,t)$ in OPO
(a), and in PCOPO  with PC affecting the pump (b), both $0.1\%$ below the corresponding threshold.
Space (ordinate axis) and time (abscissa) are scaled with diffraction length and cavity decay
as mentioned in the text.  \label{fig2}}
\end{figure}

Before to proceed to the analytical evolution of the correlation we show here an interesting  aspect
of the signal fluctuations as obtained by numerical simulations of Eqs.~(\ref{Eq:lang}).  Below
threshold noisy precursors (quantum images) dominate the dynamics as shown  in Fig.~\ref{fig2}a: the
``preferred" spatial periodicity corresponds to the critical wave-number. In the OPO the  phase of
this noisy pattern is not fixed and it diffuses in space in the $x$ direction,
being dominated by the Goldstone mode, as discussed in Ref.~\cite{2000ZambriniPRA}.
A different behavior
appears when the  PC breaks the translational symmetry  leading to a modulation of the
pump ($M_0\neq0$, as in Fig.~\ref{fig2}b).
 Then the noisy pattern  in the signal appears to be spatially
locked and there are two $\pi$-dephased modulated modes dominating the pattern. 
The consequences in terms of correlations are discussed in the following sections.

\subsection{Few mode dynamics}
\label{sec:fewmodes}

From Fig. \ref{fig1} we can now see that the assumption
$\tilde{A}_{0}^{s}(x)\approx\tilde{A}_{0}^{s}(0)+\tilde{A}_{0}^{s}(k_{p})
e^{ik_{p}x}+\tilde{A}_{0}^{s}(-k_{p})e^{-ik_{p}x}$ [Eqs. (5-7)] is well justified as it
takes into account the most relevant modes for the pump field:
higher order harmonics at $\pm 2k_p$ are much
smaller than the ones at $\pm k_p$, allowing  neglection of the contribution of
terms with $|k|>k_{p}$, i.e. modulations below the PC wavelength. On the other hand, the introduction of the modulation does not have any
effect on the signal, whose main components are always the modes at the critical wave-number
$k_{c}=k_{p}/2$. Then we restrict our analysis to the modes
$k=\pm k_{p}/2$ in Eq.~(\ref{Eq:enFourier}).
 Within this
assumption, we reduce the study of the PCOPO dynamics below threshold to four coupled operators
equations that in  the frequency domain read
\begin{eqnarray}
L\vec{a}_{1}=\sqrt{\frac{2}{\gamma}}\vec{a}_{1}^{\mathrm{in}}
\end{eqnarray}
where
\begin{eqnarray}
\vec{a}_{1} & = & \left(a_{1}(k_c),a_{1}(-k_c),a_{1}^{\dagger}(-k_c),a_{1}^{\dagger}(k_c)\right)^{\top},\\
\end{eqnarray}
and  $a_{1}(k_c,\omega)$ and  $ a_{1}^{\dagger}(k_c,-\omega)$ are
denoted as $a_{1}(k_c)$ and  $ a_{1}^{\dagger}(k_c)$.
The vector $ \vec{a}_{1}^{\mathrm{in}}$ is expressed in a similar manner. The matrix $L$ is
\begin{equation}
\label{Eq:matL}
L=\left(\begin{array}{cccc}
1-i\omega & \frac{M_{1}}{2} & -S & -\kappa S \\
-\frac{M_{1}}{2} & 1-i\omega & \kappa S & -S \\
-S^{*} & \kappa^{*}S^{*} & 1-i\omega & -\frac{M_{1}}{2}\\
-\kappa^{*}S^{*} & -S^{*} & \frac{M_{1}}{2} & 1-i\omega\end{array}\right).
\end{equation}
From Eqs.~(\ref{eq:stationaryk}) and (\ref{eq:stationaryk0}),  with $k_p=2k_c$,
 we obtain
\begin{equation}
\label{eq:S}
{A}_{0}^{s}(0) =S=\frac{E\left(1-i2\Delta_{1}\right)}{1-i2\Delta_{1}+M_{0}^{2}/2},
\end{equation}
and ${A}_{0}^{s}(\pm k_p)=\pm \kappa S$,  with
\begin{equation}
\label{eq:kappa}
\kappa=\frac{-M_{0}/2}{1-i2\Delta_{1}}.
\end{equation}
The output fields are obtained from
the input-output formalism
$\vec{A}^{\mathrm{out}}=\sqrt{2\gamma}\vec{A}_{1}-\vec{A}_{1}^{\mathrm{in}}$~\cite{1984ColletPRA}, and
their dynamics (in the frequency domain $\omega$) is governed by
\begin{equation}
\label{Eq:In-out}
\vec{a}^{\mathrm{out}}=\left(2L^{-1}-\mathbb{I}\right)\vec{a}_{1}^{\mathrm{in}},
\end{equation}
where $\mathbb{I}$ is the $4\times4$ identity. In the following  we will concentrate on spatial
quantum effects in the signal field. Therefore,  we can omit without ambiguity the index $1$. To calculate different correlations in the output
variables given the input ones, we need to obtain the inverse of the matrix $L$ 
(the expression for this inverse is given in
App.~\ref{sec:InverseL}). This formalism is at the basis
of the  analytical  quantum correlation for the output fields discussed in the following sections.

\section{Intensity correlations and the parametric threshold in the  presence of a PC }
\label{Sec:threshold}

\begin{figure}
\begin{tabular}{c}
\hspace{-1.1cm}\includegraphics[ width=7cm]{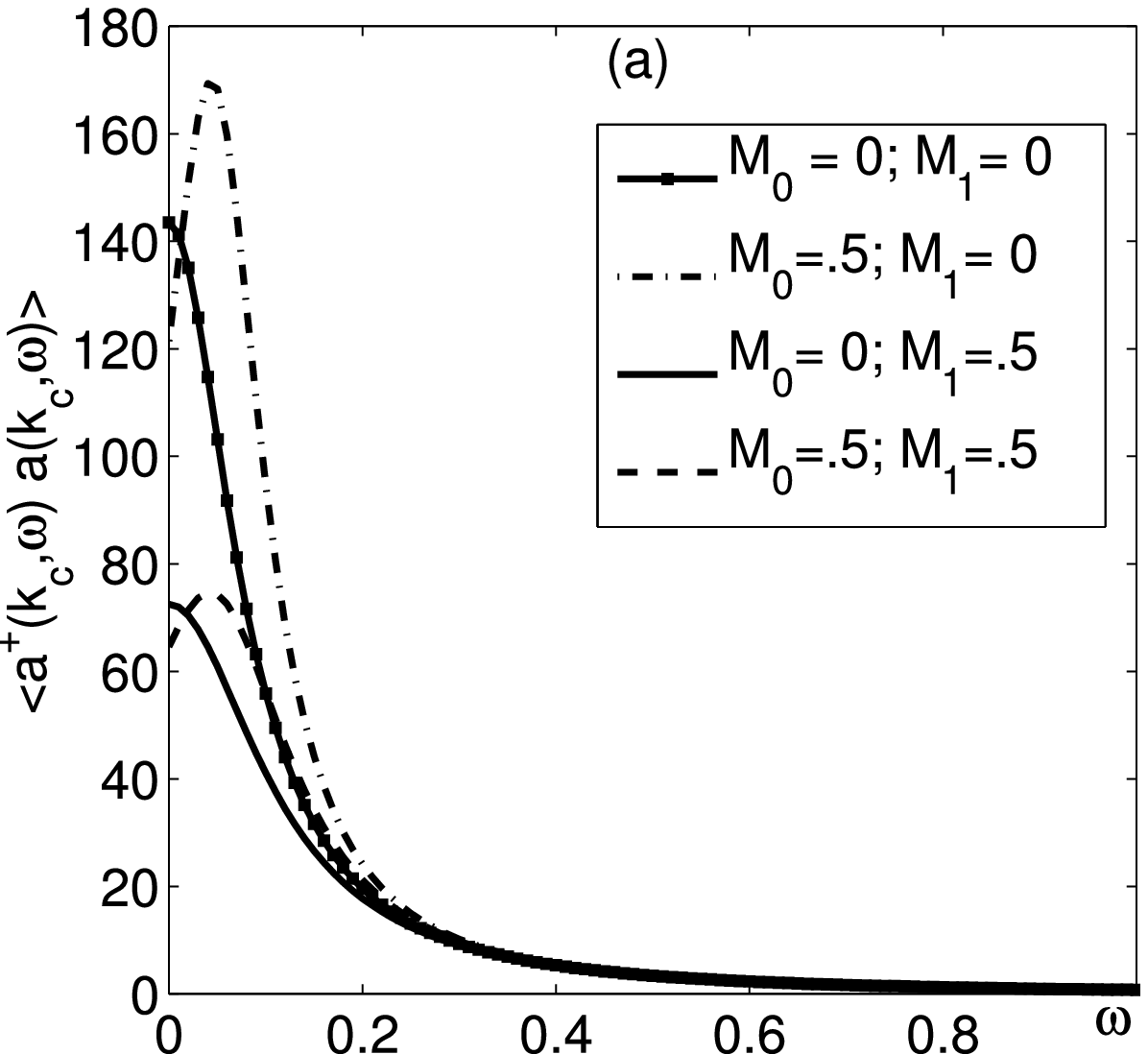} \\
\includegraphics[trim =20mm 1mm 1mm 1mm, clip, width=8cm]{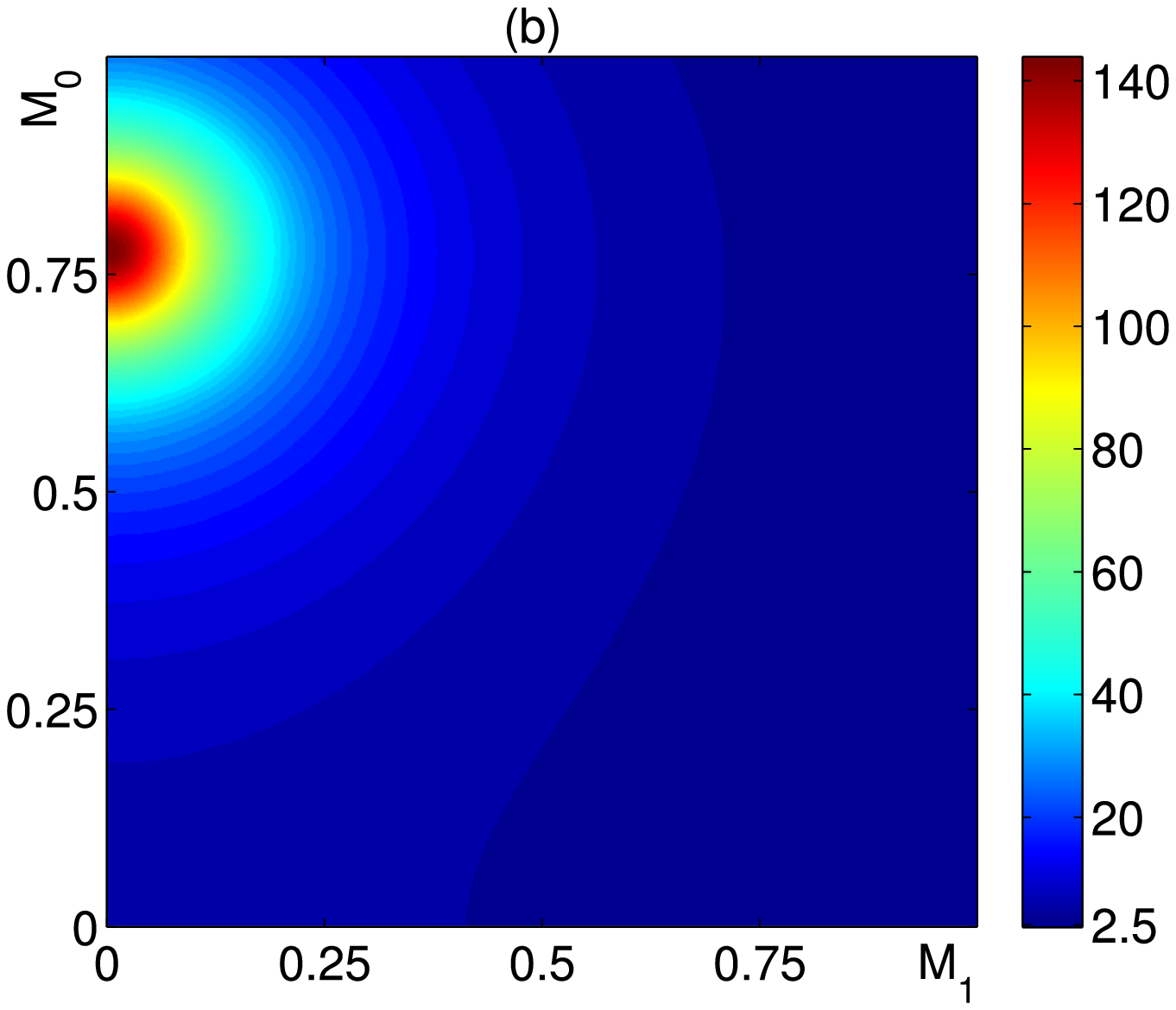} \vspace{-0.25cm}\\
\hspace{1.1cm}\includegraphics[trim =20mm 1mm 1mm 1mm, clip, width=8cm]{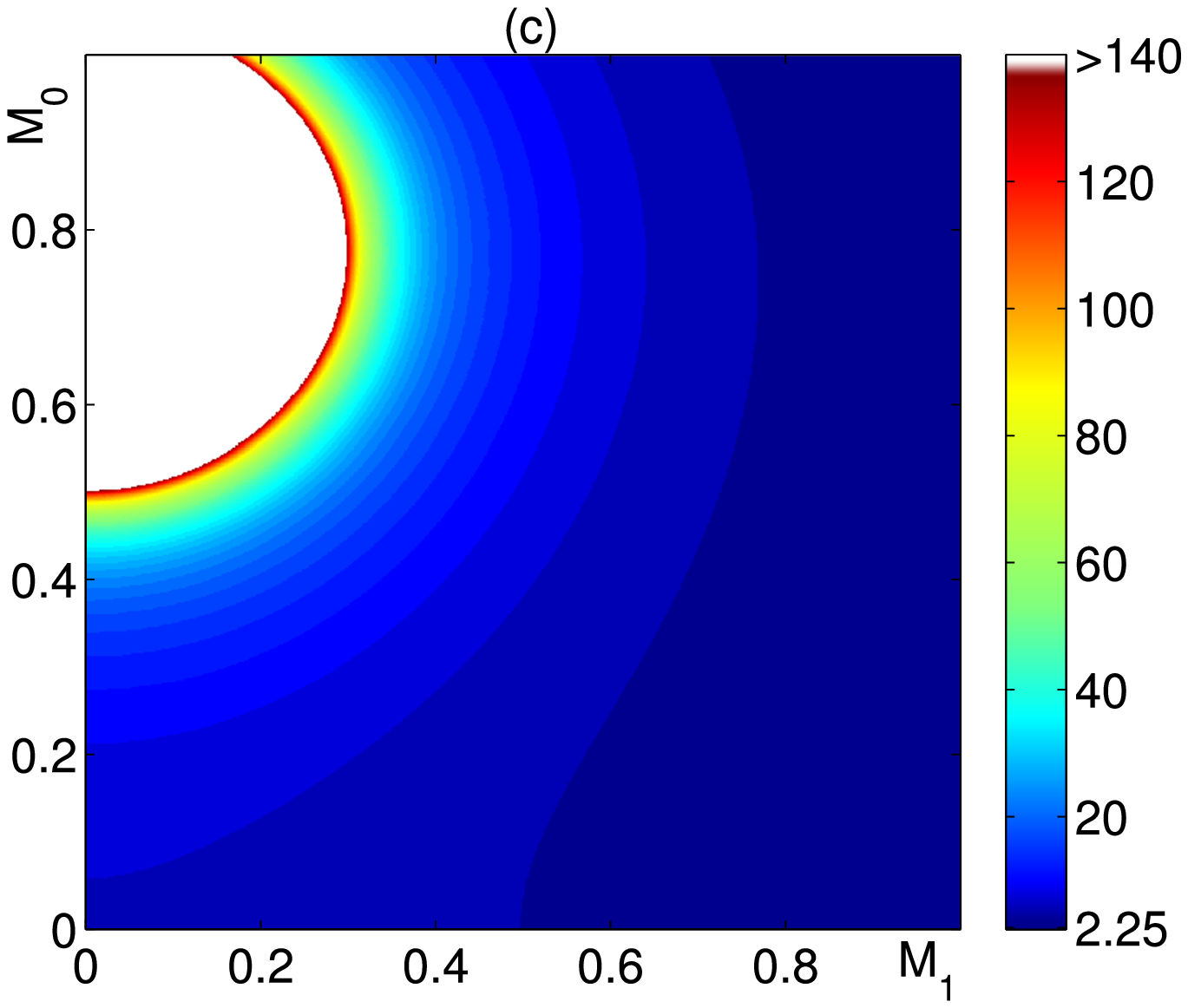}  \vspace{-0.5cm}
\end{tabular}
\caption{(Color on line) {\it Intensity in frequency and  time domain for fixed values of the external pump, and different configurations of the PC}. In (a) we show the intensity for $E=0.92$ in frequency domain for four different configurations. In (b) we show it in the time domain for the same pump,  and for different combinations of the parameters $M_0, M_1$. In this case, the threshold is not reached for any configuration. In (c) we show that, if the external pump is increased to $E=0.93$,  the intensity tends to infinity for the combinations of the parameters inside the white circle.  \label{fig3}}
\end{figure}
Let us discuss the quantum effects of introducing a PC in an OPO below threshold, starting
from the intensity $\left\langle n^{\mathrm{out}}\right\rangle =\left\langle
a^{\mathrm{out},\dagger}\,a^{\mathrm{out}}\right\rangle $ of the  most intense mode ($k_c=k_p/2$).
The analytical expressions of second order moments are given in Appendix~\ref{sec:Correlationsw}.
In Fig.~\ref{fig3}a we represent the spectral intensity, $\left\langle
n^{\mathrm{out}}(k_c,\omega)\right\rangle $, whose analytical expression is given in Eq.~\ref{eq:espectI}, for
different configurations of the PC. The effect of the PC is immediately recognized on the intensity
spectrum: not only the intensity at each frequency component can be largely increased/decreased with
respect to the case of the OPO, but also the maximum of the spectrum can appear shifted 
(away from
$\omega=0$)
when the pump
detuning is modulated, that is for $M_0\ne0$.

The (stationary) intensity is obtained after some standard but cumbersome
calculations  (Appendix~\ref{sec:Correlationsw}) and reads:
\begin{align}
\label{Eq:Intensity}
& \left\langle a^{\mathrm{out},\dagger}(k_c,t)a^{\mathrm{out}}(k_c,t)\right\rangle  =  \\
& -4|S|^{2}\left(4|S|^{2}|1+\kappa^{2}|^2 - \left(1+|\kappa|^{2}\right)\left(4+M_{1}^{2}\right)\right)/\sigma,\nonumber
\end{align}
where the denominator $\sigma$ is
\begin{align}
\label{Eq:denominator}
 \sigma & = 16|S|^{4}|1+\kappa^{2}|^2-8|S|^{2}\left(1+|\kappa|^{2}\right)\left(4+M_{1}^{2}\right)\nonumber\\
& +\left(4+M_{1}^{2}\right)^{2}.
\end{align}
with $S$ and $\kappa$ given in Eqs.~(\ref{eq:S}) and (\ref{eq:kappa}), respectively.

It is easy to see
that this expression reduces to the one given in~\cite{1997GattiPRA} in absence of the PC, when
$M_0=M_1= 0 $, i.e. $E^{2}/(1-E^{2})$.   In a previous work we showed that the numerical results of the full
model~(\ref{Eq:lang}) are in a good agreement with this analytical expression
(see Fig.~2 of
Ref.~\cite{2011DeCastro}), thus
justifying the assumptions described above.

In general, for the PCOPO, we see that both pump and signal modulations can modify the signal emission
at the most intense spatial mode $k_c$.
In Fig.~\ref{fig3}b we represent the intensity obtained from Eq.~(\ref{Eq:Intensity}) as a function
of $M_0$ and $M_1$, for a fixed value  of the external  pump, $E=0.92$. The intensity can be controlled
through the PC and a strong effect is found for small modulations of the signal detuning
$M_1$ and certain values of the modulation of the pump $M_0$ ( $M_1\simeq 0$
and $M_0\simeq 0.8$ in Fig.~\ref{fig3}b).  While for this value  of the external pump
$E$ the intensity remains finite for every combination of $M_1$ and $M_0$,  
this is not necessarily true when the pump is increased.

Indeed, for larger pump $E$
the average intensity in the PCOPO increases and, eventually, the
threshold is reached. When approaching the parametric down-conversion threshold (coinciding with the
spatial instability one) our approximation would fail and this is at the basis of the divergences
appearing in Eq.~(\ref{Eq:Intensity}).  In
Fig.~\ref{fig3}c we represent the intensity for an external pump $E=0.93$ and different
configurations of the PC. At the edges of the white circle the intensity tends rapidly to infinity.
Then, the  presence of the PC has reduced the instability associated with the threshold to a value
smaller than $E=0.93$ for the points inside the white the circle,
while in absence of a PC (for $M_0=M_1=0,\Delta_0=0$) the threshold is reached at $E=1$.

This is not the only possible scenario. As a matter of fact band-gaps are known to reduce spontaneous
emission~\cite{2008Joannopoulos,1987YablonovitchPRL,1987JohnPRL}. Therefore,
for our choice $k_{\mathrm{pc}}=2k_c$, the PC is expected to cause
reduction of the  fluctuations that would lead to instability, inhibiting it. Indeed, the threshold can be either
lowered or raised by means of the PC, as discussed in~\cite{2011DeCastro}.
As the signal wavenumber $k_c$ is in the band-gap~\cite{2004GomilaPRL, 2005GomilaPRE},
inhibition of
the spatial instability  (lowering the threshold) is expected. On the other hand, the pump wavenumber
is  $2k_c$, and  therefore the PC introduces in the system exactly the wavenumber at which the
instability process should occur in the pump, imprinting a nonlinear structure that favors the
instability. Then a raise of the threshold can be expected. We find that these two competing mechanisms
can increase or decrease the threshold depending on the relative values of the different amplitudes of
the spatial detuning   $\Delta_0(x)$ and $\Delta_1(x)$.

For $M_0=0$  we can
easily obtain  an expression for the increase of the threshold with $M_1$. In this case, the
divergence of the intensity, when
Eq.~(\ref{Eq:denominator}) vanishes, leads to the threshold expression
\begin{align}
\label{Eq:thresh}
 E_{\mathrm{thr}}=\sqrt{1+(M_1/2)^2},
\end{align}  in accordance with the results given
in~\cite{2005GomilaPRE}. 
This curve is indeed the black curve representing the threshold in
Fig.~\ref{fig4}c. Moreover, in Fig.~\ref{fig4}b, the black curve represents the thresholds for
$M_0\ne0$ and $M_1=0$, thus showing that the threshold can be either raised or lowered with $M_0$.
Finally, notice that if we increase further $E$ eventually reaching $E=1$, the intensity remains finite
in some regions, in accordance with the fact that the threshold is increased for some configurations of
the PC.

\section{Quantum correlations in the presence of a PC}
\label{Sec:Qcorr}

As clear from Fig.~\ref{fig2} there are profound effects on the fluctuations of the signal field  of an
OPO when including the PC modulation. Changes in the correlations between the most intense modes $
a^{\mathrm{out}}(k_p/2)$ and $a^{\mathrm{out}}(-k_p/2)$ are then expected and in the following we consider quantum
effects such as squeezing, entanglement and twin beams correlations.

\subsection{Squeezing in the presence of PC}

Optical parametric oscillators are well-known as sources of spatial squeezing as
mentioned in the Introduction~\cite{1997GattiPRA,2003ZambriniPRA}.
Two mode squeezing appears in quadratures of the
superposition of two opposite signal modes:
\begin{align}
\label{x's}
& \Sigma_{\theta\varphi}\left(k_c,-k_c\right)  =\\
&  \left[a^{\mathrm{out}}\left(k_c\right)+e^{i\varphi}a^{\mathrm{out}}\left(-k_c\right)\right]e^{i\theta}
+\mbox{h.c.},\nonumber\end{align}
where $\theta$ is the quadrature angle and $\varphi$ is a relative phase or superposition
angle. This is equivalent to a sum of (position) quadratures
$\Sigma_{\theta\varphi}\left(k_c,-k_c\right)= \hat{x}_1 +\hat{x}_2$, where:
\begin{eqnarray}
\label{eq:xi}
& \hat{x}_1 = a^{\mathrm{out}}\left(k_c\right)e^{i\theta}+a^{\mathrm{out},\dagger}\left(k_c\right)e^{-i\theta}\\
&\hat{x}_2 = a^{\mathrm{out}}\left(-k_c\right)e^{i(\theta+\varphi)}+a^{\mathrm{out},\dagger}\left(-k_c\right)e^{-i(\theta+\varphi)} \nonumber
\end{eqnarray}
are the quadratures corresponding to the mode at $k_c$ and $-k_c$, respectively.

The variance $\Delta^2 \Sigma_{\theta\varphi}\left(k_c,-k_c\right)$  can be obtained in
terms of the operator moments describing the fields
out  of the  cavity  as given in  App.~\ref{sec:Correlationsw}. We remind that
the first order moments of the signal field in all spatial modes vanish below
threshold, so that $\Delta^2 \Sigma_{\theta\varphi}$ is obtained from the second order
moments of field operators.
%
\begin{figure}[t]
\begin{tabular}{cc}
\includegraphics[trim =25mm 1mm 0mm 5mm, clip, width=5.cm]{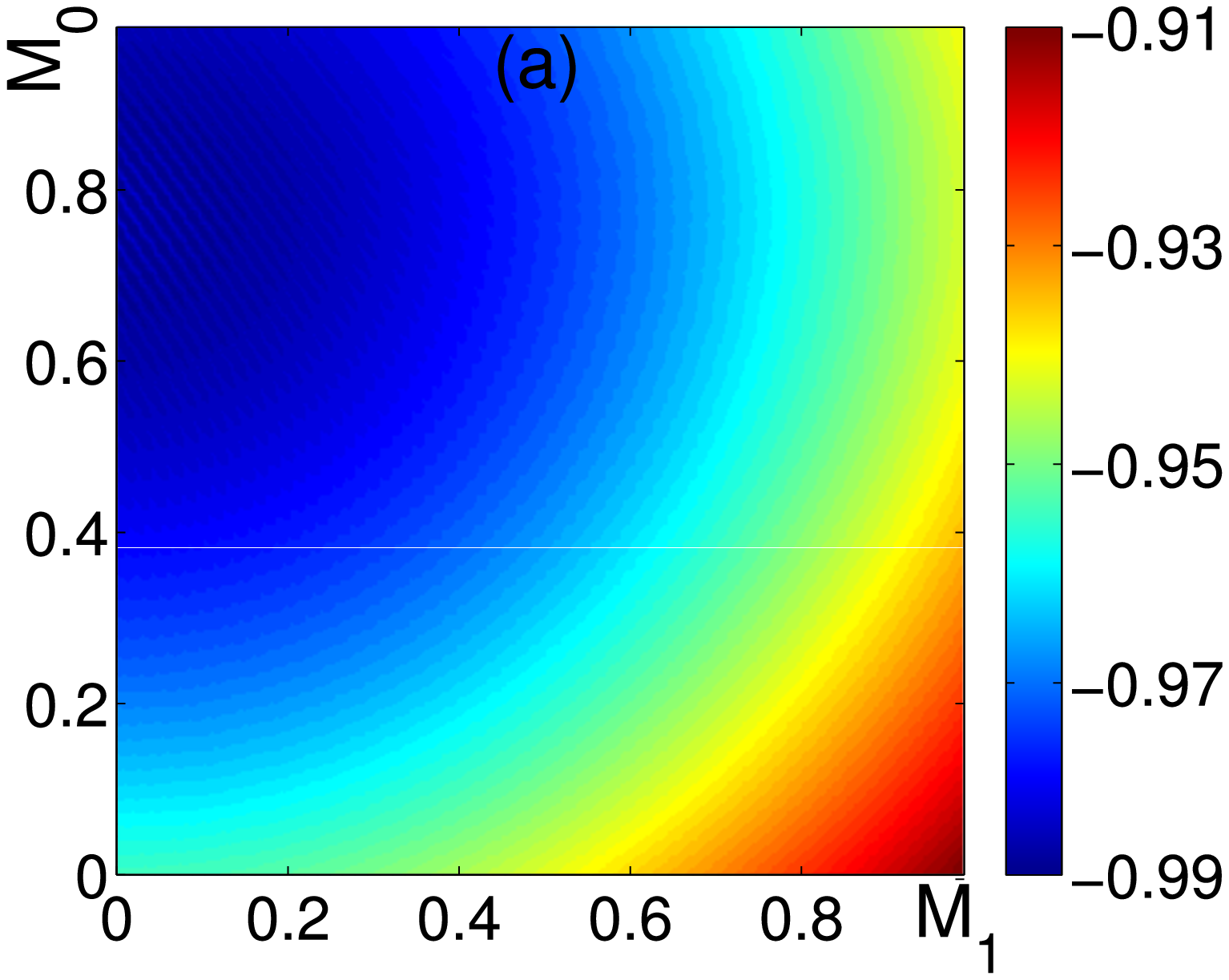} & \hspace{-.75cm}\includegraphics[trim =5mm 1mm 0mm 5mm, clip, width=4.1cm]{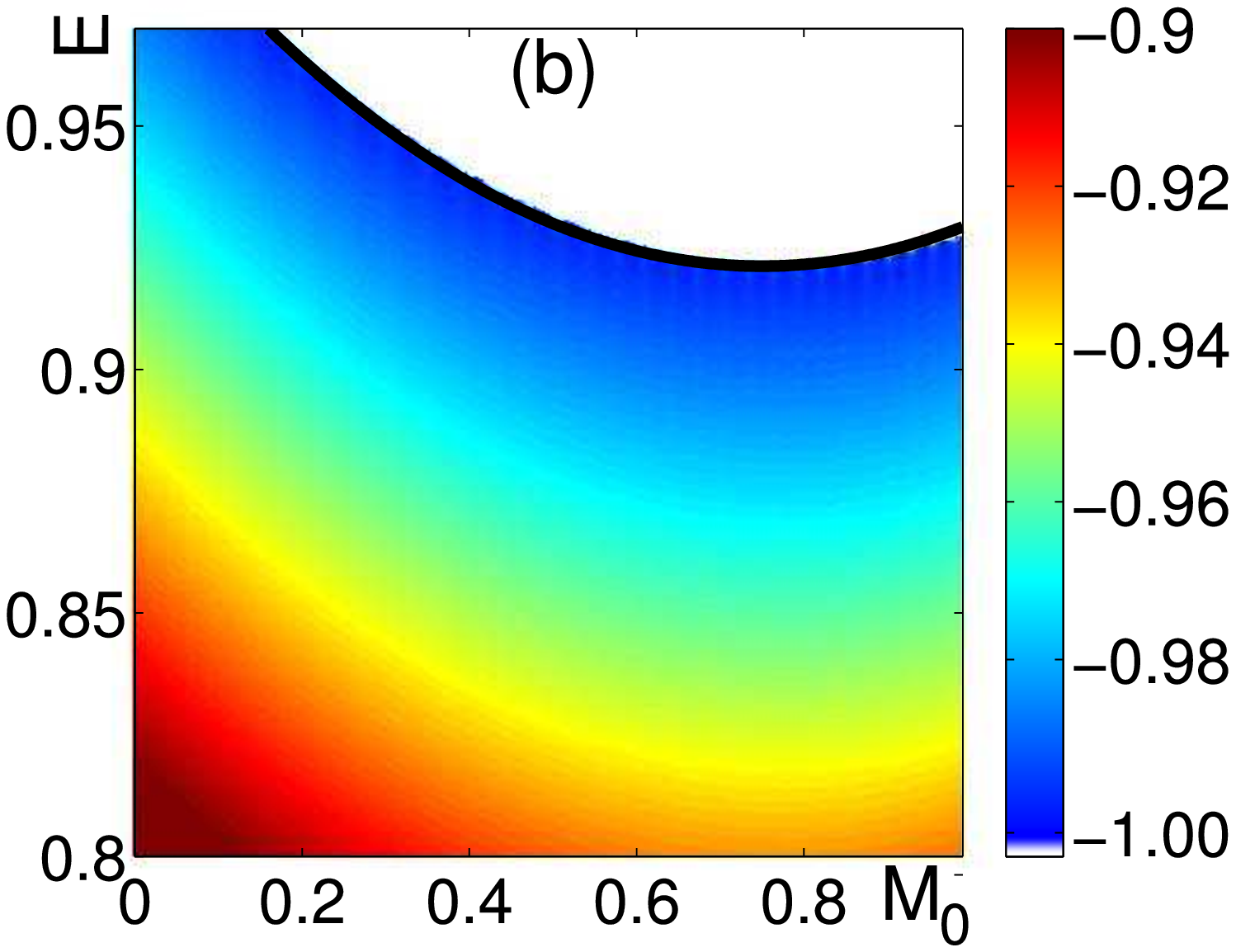}\\
\hspace{-.95cm}\includegraphics[trim =10mm 1mm 0mm 0mm, clip, width=4.2cm]{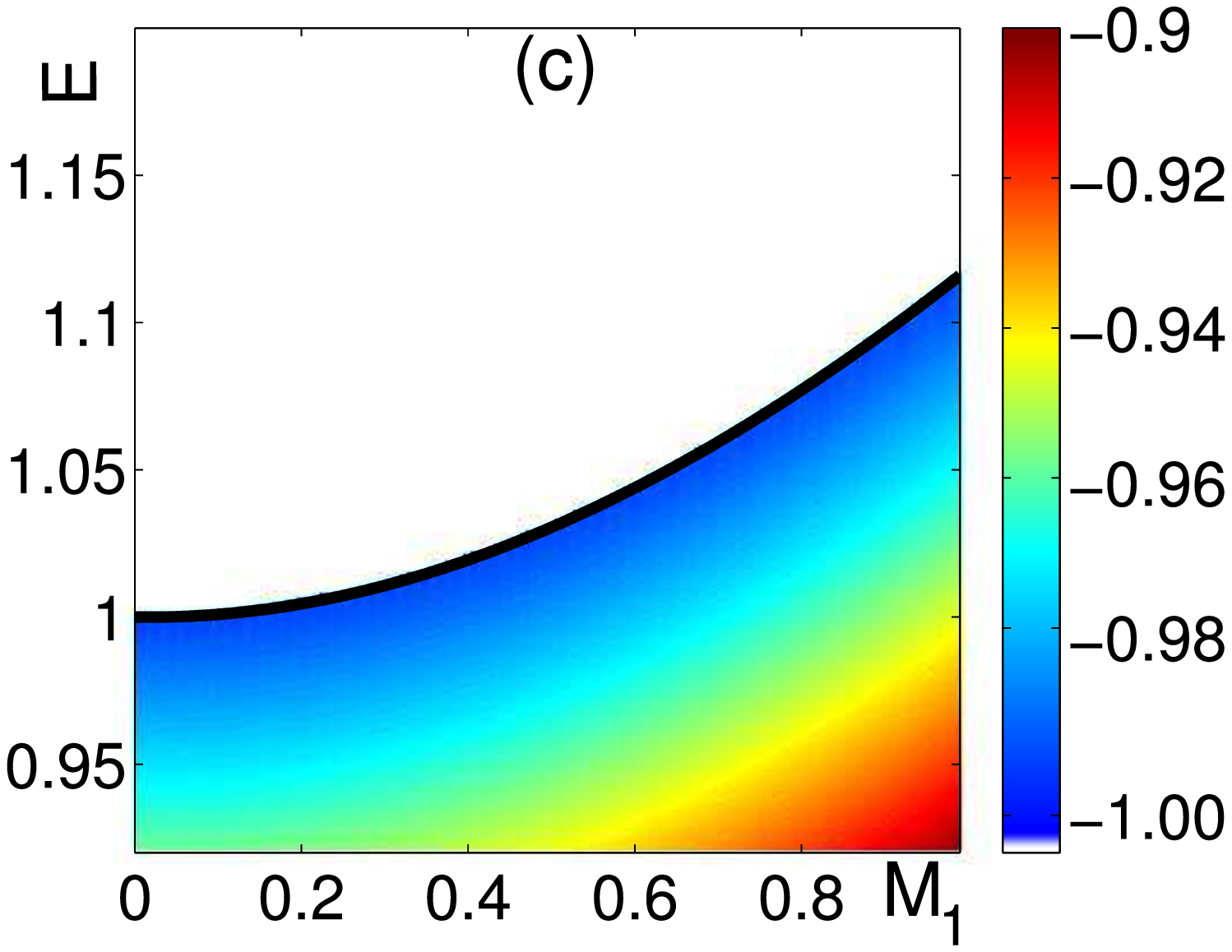} &
\hspace{-1.25cm}\includegraphics[trim =5mm 1mm 0mm 0mm, clip, width=3.8cm]{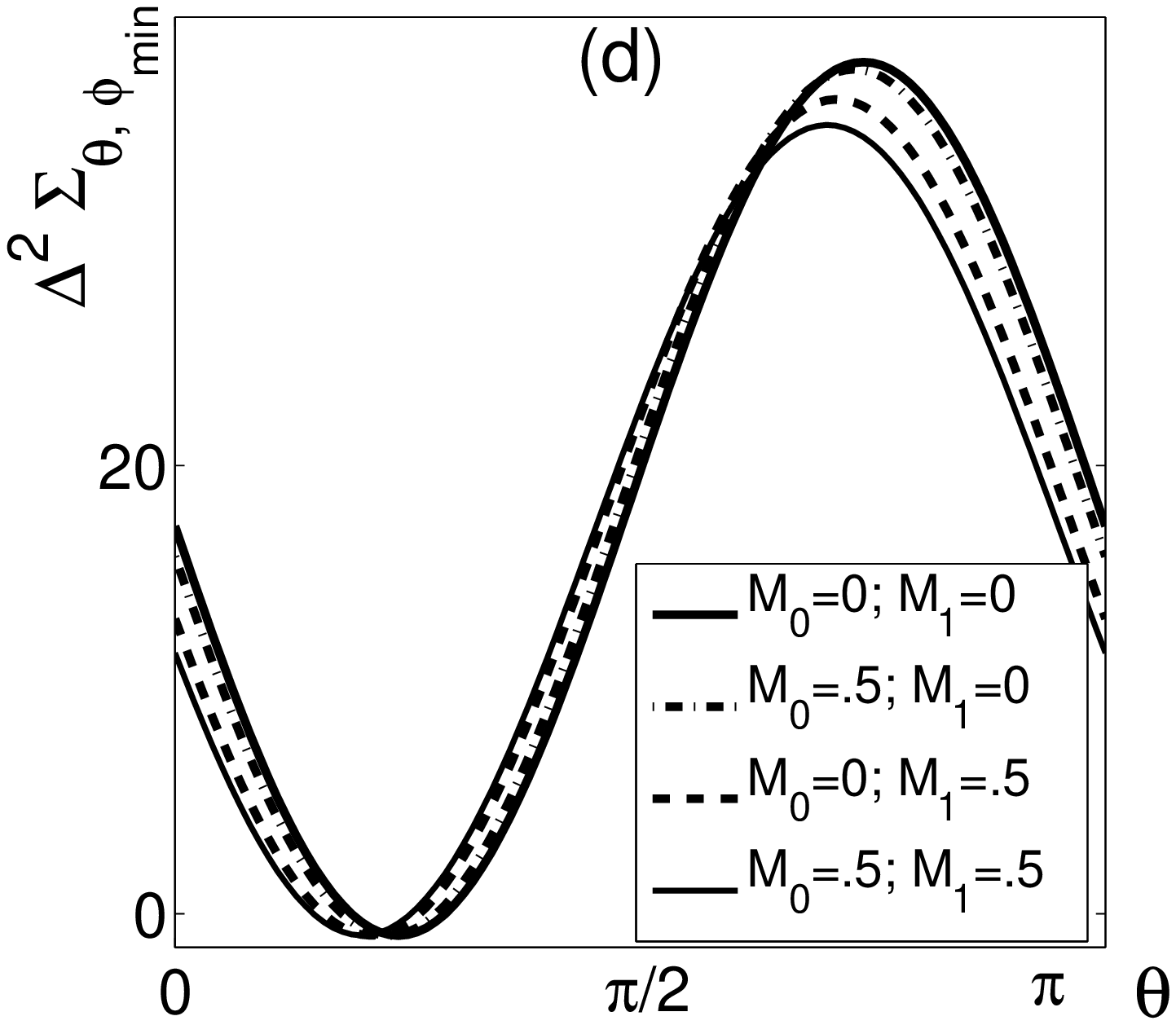}\vspace{-0.25cm}\\
\end{tabular}
\caption{(Color online) {\it Minimum value of the variance for different configurations of the PC}. (a) Minimum of the variance for an external pump below threshold, $E=0.92$. (b) The variance for every $\theta$ for $E$ at a $5\%$ below threshold. Here, $\phi$ is fixed to the value at which the minimum variance  is obtained. (c) Minimum value of the variance for different combinations of $M_0$ and  $E$ below threshold,  with when $M_1=0$. (d) same when $M_1\ne0$ and $M_0=0$. This minimum is never smaller than the theoretical value in absence of PC, and it reaches this value at the corresponding threshold $E_{\mathrm{thr}}$, which in (c) and (d) is represented by the black solid curve.   \label{fig4}}
\end{figure}

From the analytical expression of the variance we can find the squeezing for different combinations of
$M_0$, $M_1$, and the pump $E$. The latter is always considered below  threshold for every combination of
$M_0$ and $M_1$. We observe that the level of squeezing reached depends not only in the pump $E$ but also
in the values that define the PC. For example, for a fixed value of the pump $E$ it
depends on $M_0$ and
$M_1$, as represented in Fig.~\ref{fig4}a. On the other hand, this level also depends on $E$ and $M_0$ for
fixed values of $M_1$ (see Fig.~\ref{fig4}b), or   $M_1$ for fixed values of $M_0$ (see Fig.~\ref{fig4}c).
The level of squeezing depends on the distance to the threshold, being enhanced when the system 
gets
closer to the instability threshold. In an OPO, one can drive the system towards the instability
by increasing $E$. A PCOPO permits one to control the value of the threshold through the PC, thus allowing
one to control the squeezing by modifying $M_0$ and $M_1$.  
In Fig.~\ref{fig4}d we represent how
the variance varies with the quadrature angle $\theta$, for the a value of the superposition angle $\phi$
at which the maximum squeezing is achieved. For all four cases, we consider that the external pump is a
$5\%$ smaller than its corresponding value at threshold for each configuration. Since the distance to the
threshold is similar, the minimum variance reached for every configuration is also similar, and 
thus the same
level of squeezing is obtained for different configurations. 
\begin{figure}
\begin{tabular}{cc}
\includegraphics[trim =7mm 0mm 0mm 0mm, clip, width=4.5cm]{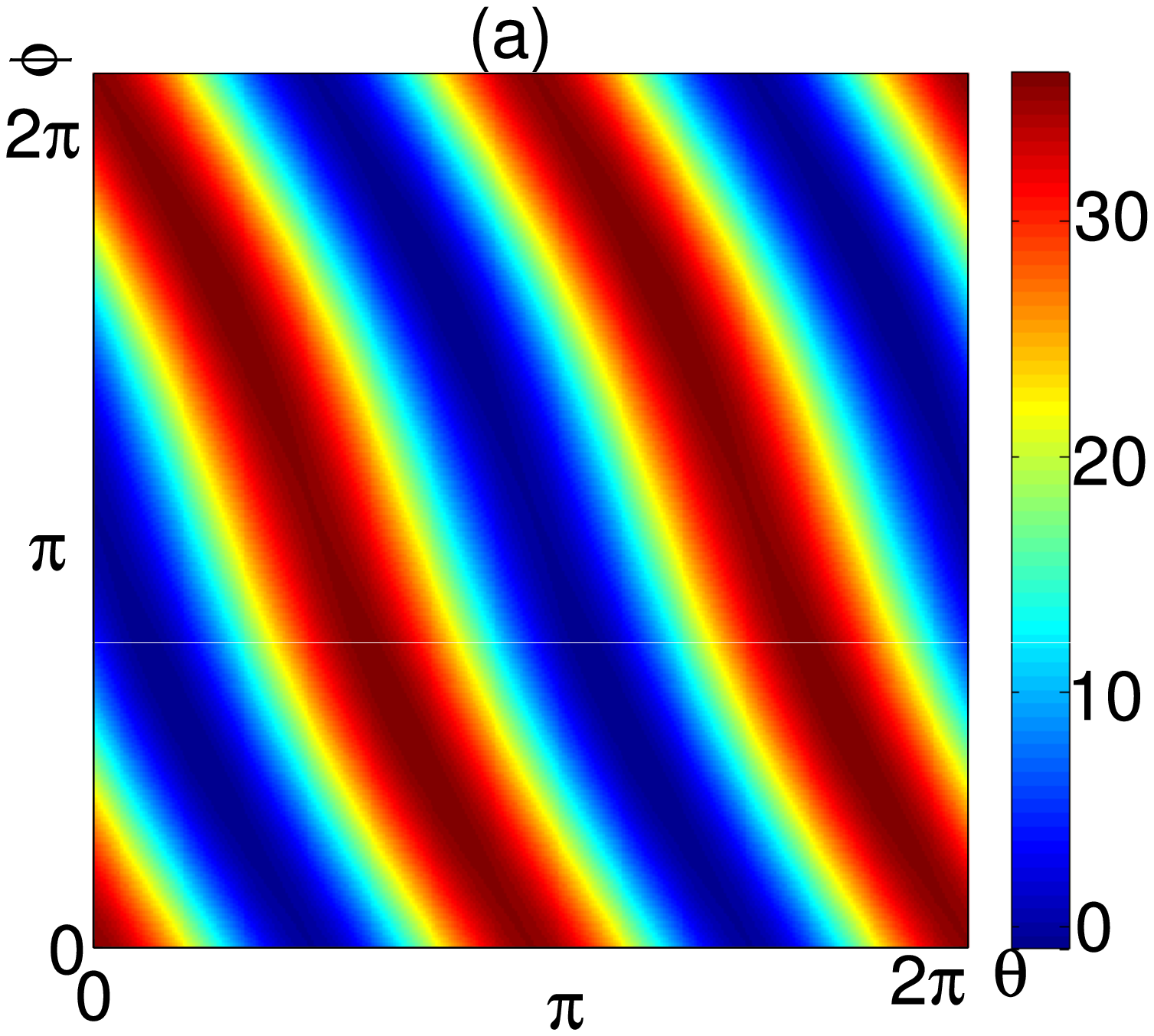}   &
\includegraphics[trim =7mm 0mm 0mm 0mm, clip, width=4.5cm]{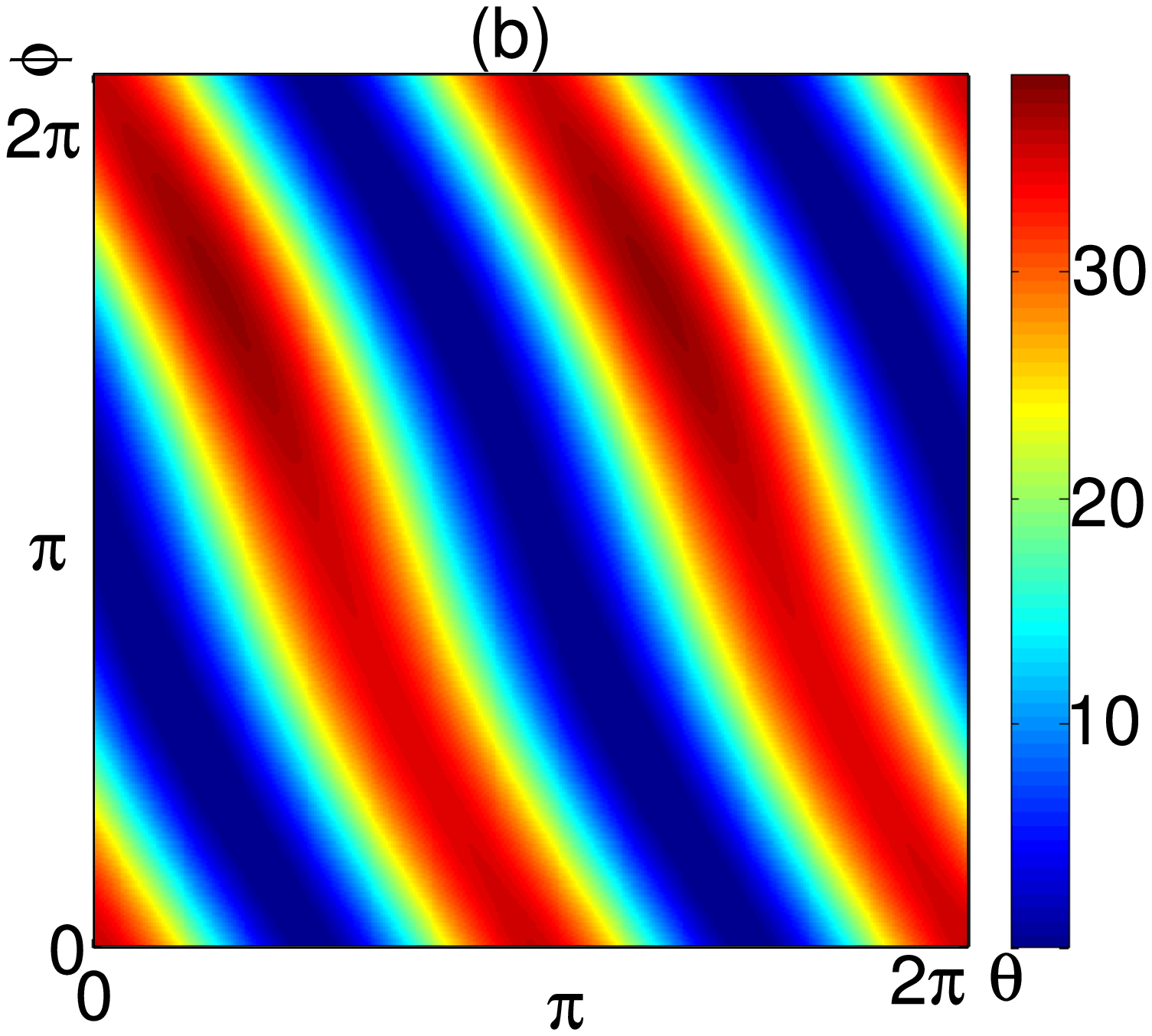}\\
 \includegraphics[trim =7mm 0mm 0mm 0mm, clip, width=4.5cm]{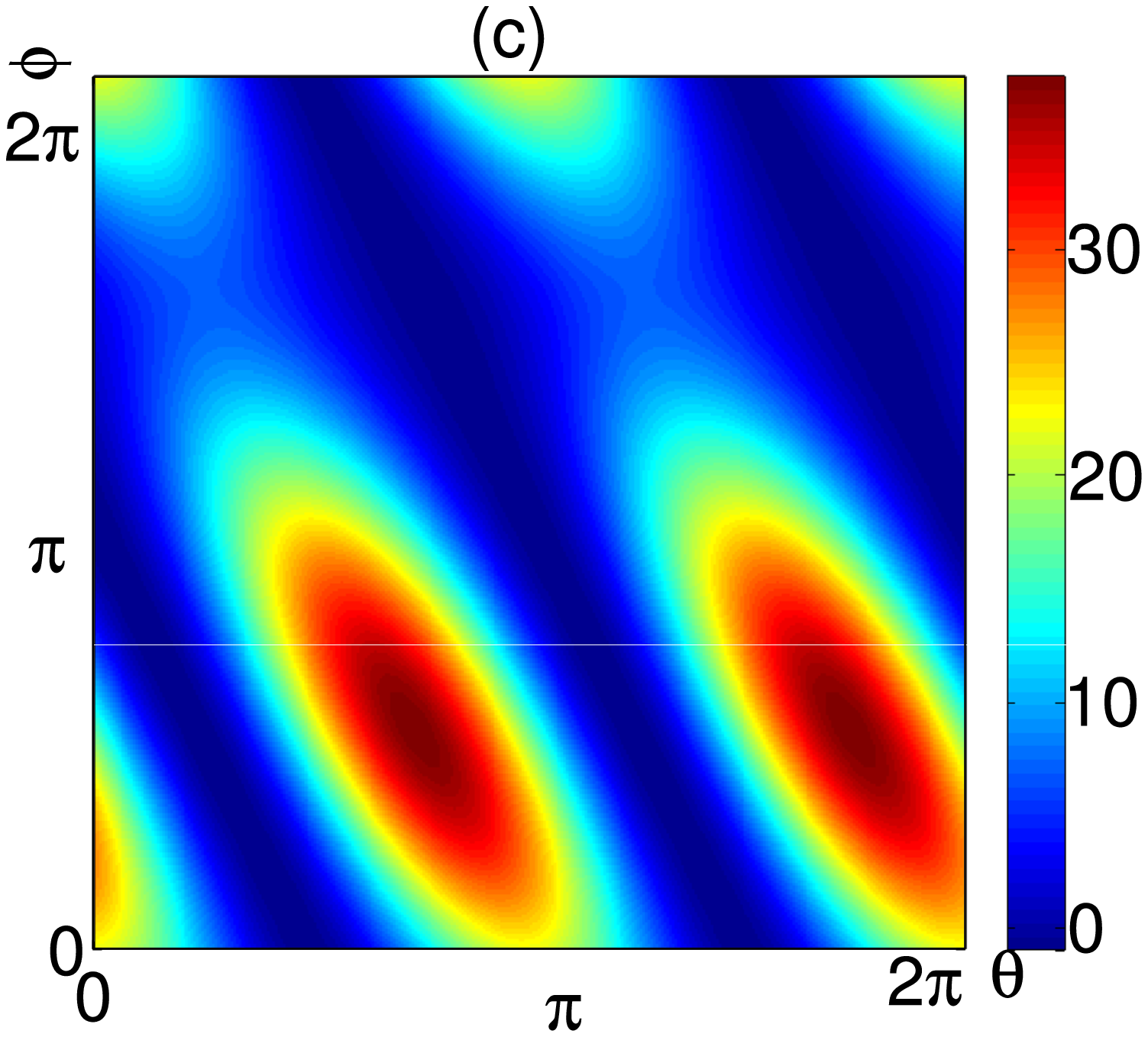}   &
\includegraphics[trim =7mm 0mm 0mm 0mm, clip, width=4.5cm]{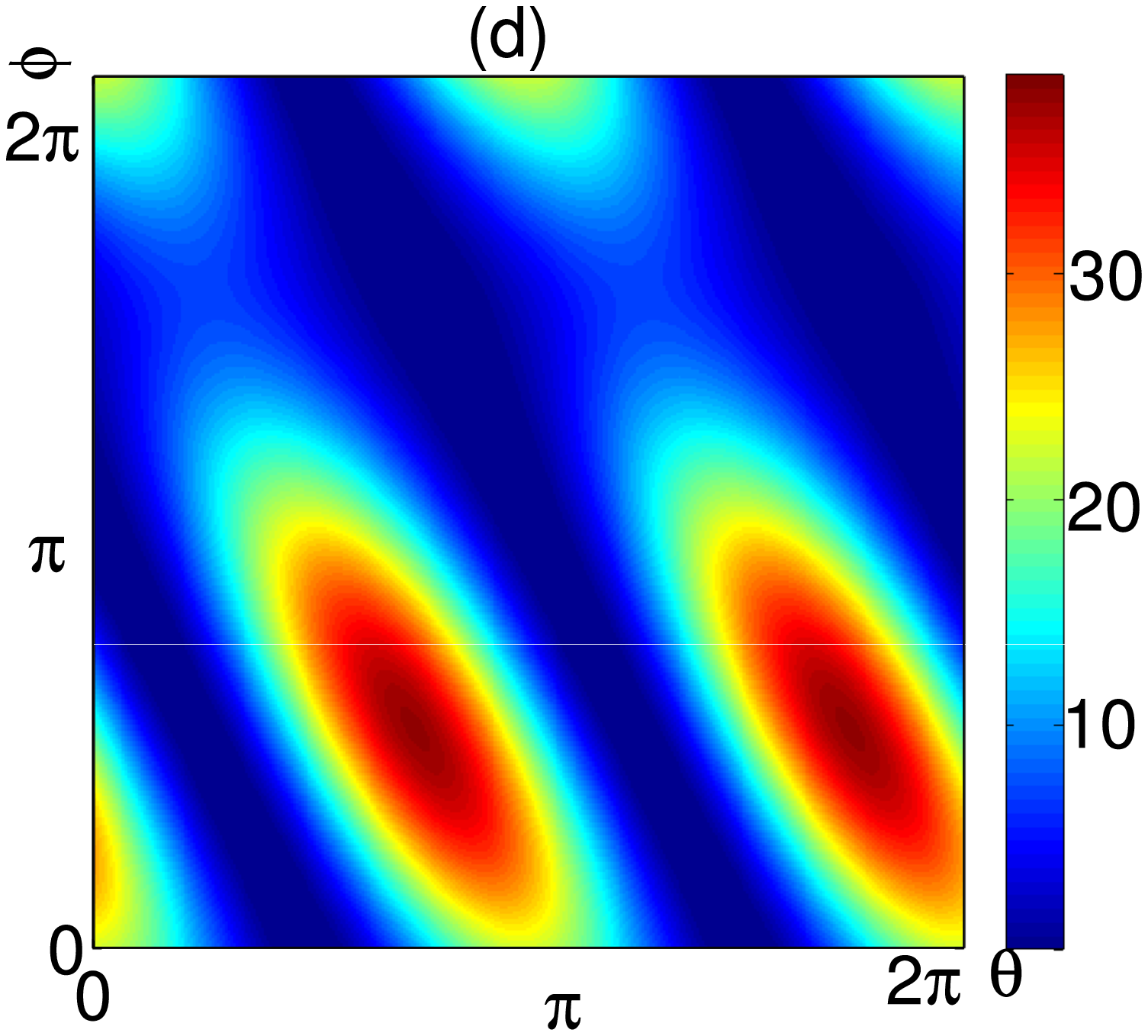}
\end{tabular}
\vspace{-0.5cm}
\caption{(Color online) {\it Variance for different configurations of the PC at all angles}. Left column contains analytical results, compared with the right column of numerical ones. All the cases are represented for a $5\%$  below threshold. (a) and (b) are for PCOPO with PC only affecting the pump ($M_0=0.5$, $M_1=0$) while (c) and (d) with PC only affecting the signal ($M_0=0$, $M_1=0.5$).  \label{fig5}}
\end{figure}
\begin{figure*}[t]
\begin{tabular}{cccc}
\includegraphics[trim =7mm 0mm 0mm 0mm, clip, width=4.5cm]{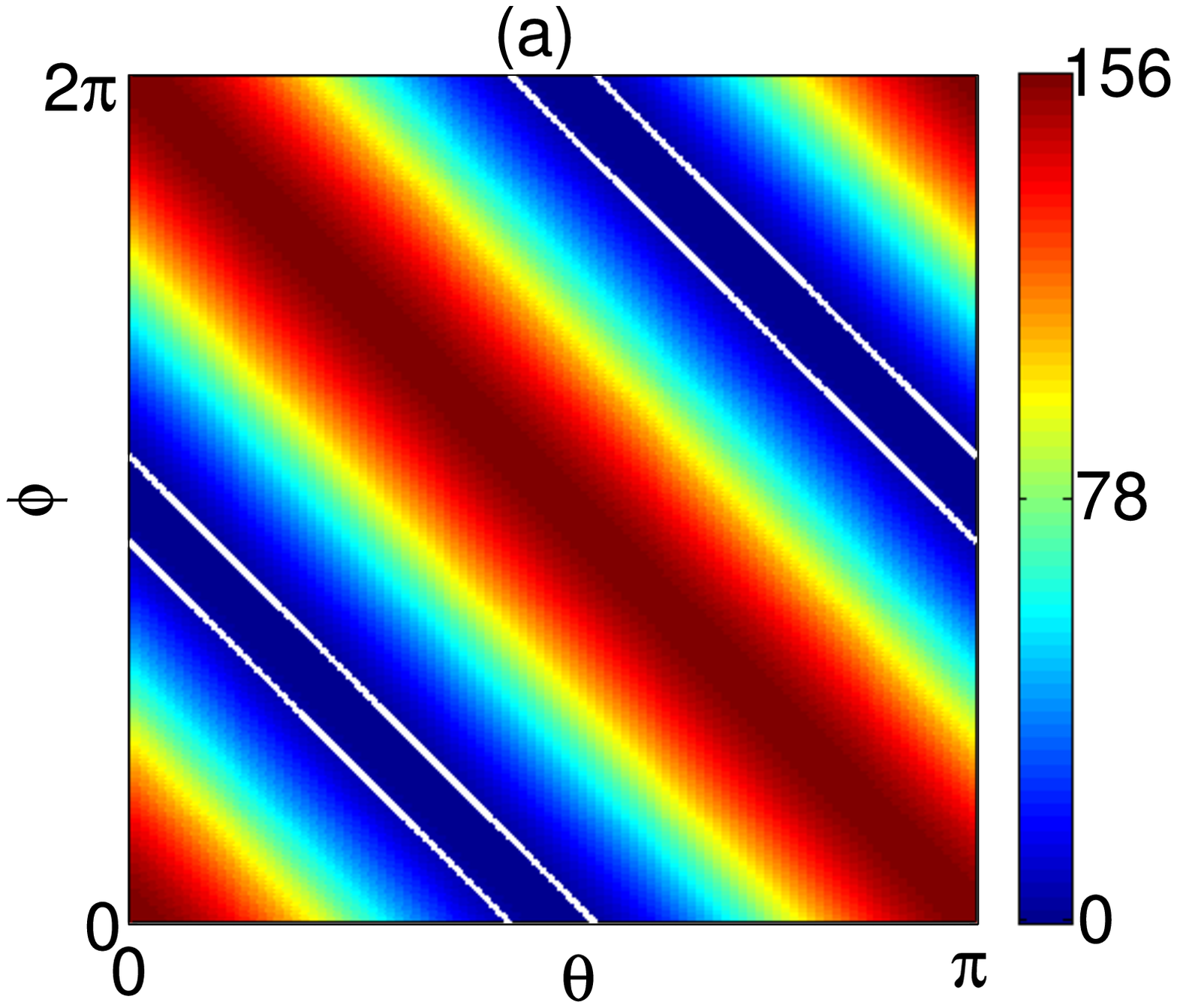} &
\hspace{-.25cm}\includegraphics[trim =7mm 0mm 0mm 0mm, clip, width=4.5cm]{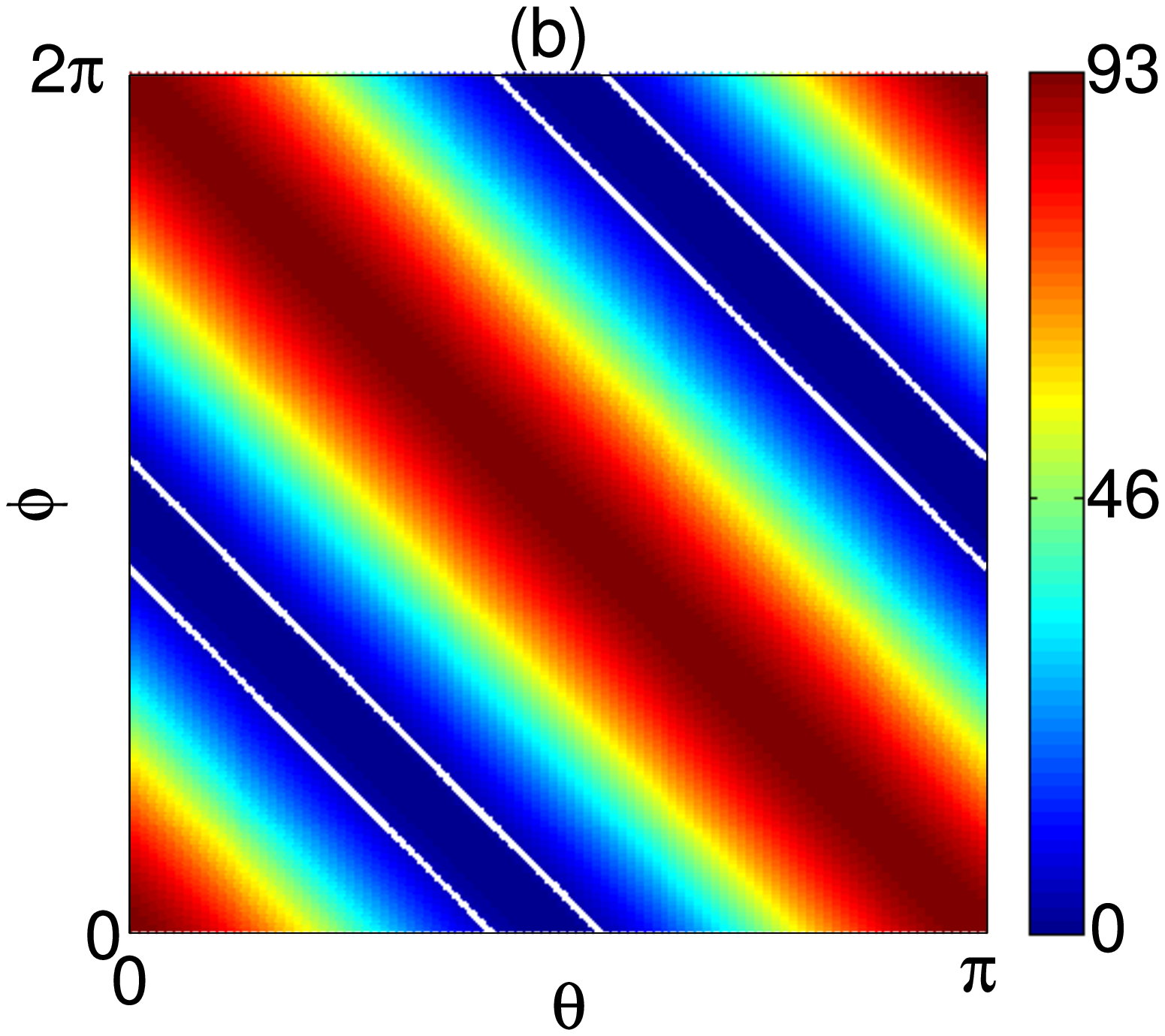} &
\hspace{-.25cm}\includegraphics[trim =7mm 0mm 0mm 0mm, clip, width=4.5cm]{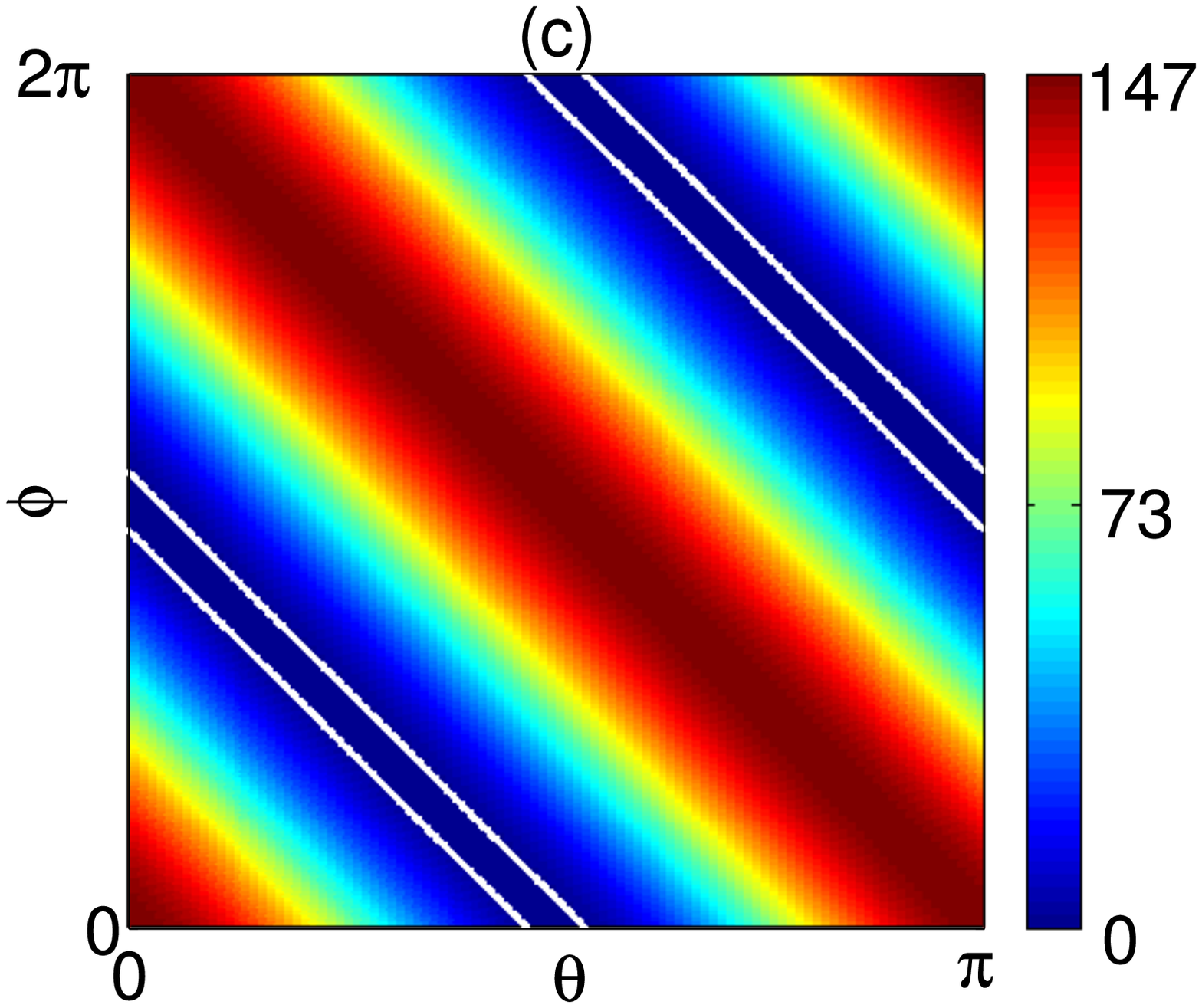} &
\hspace{-.25cm}\includegraphics[trim =7mm 0mm 0mm 0mm, clip, width=4.5cm]{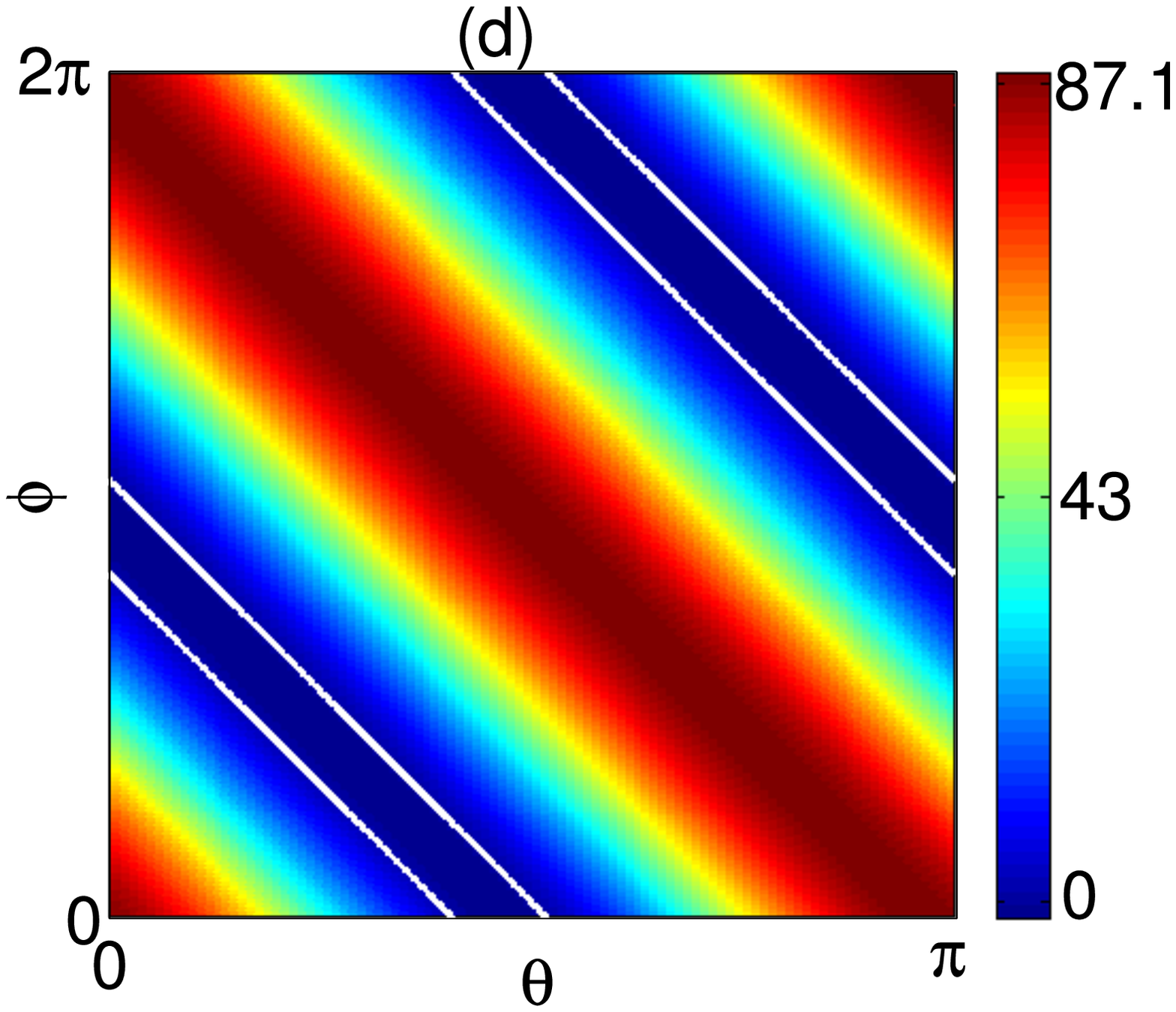} \\
\includegraphics[trim =7mm 0mm 0mm 0mm, clip, width=4.5cm]{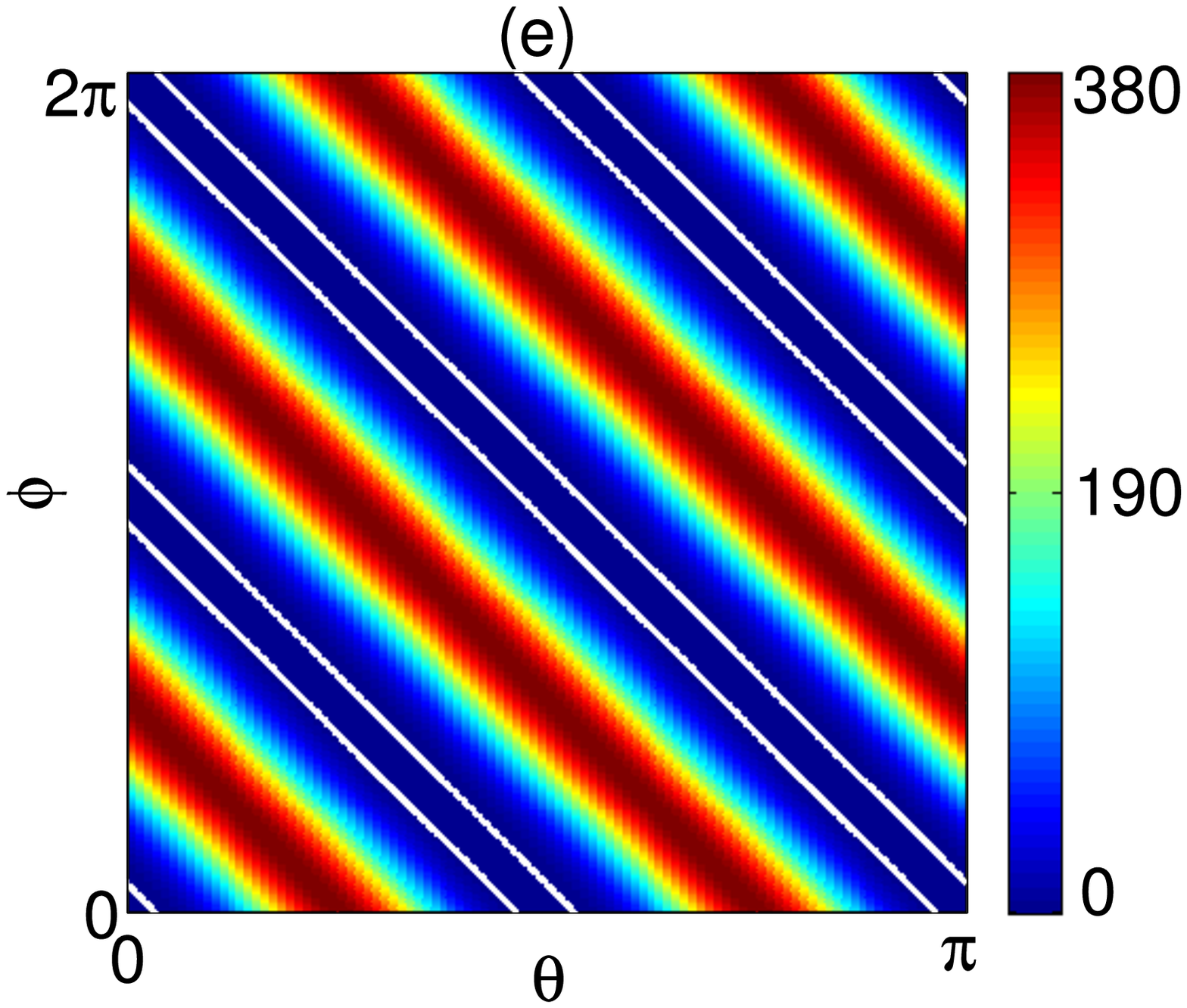} &
\hspace{-.25cm}\includegraphics[trim =7mm 0mm 0mm 0mm, clip, width=4.5cm]{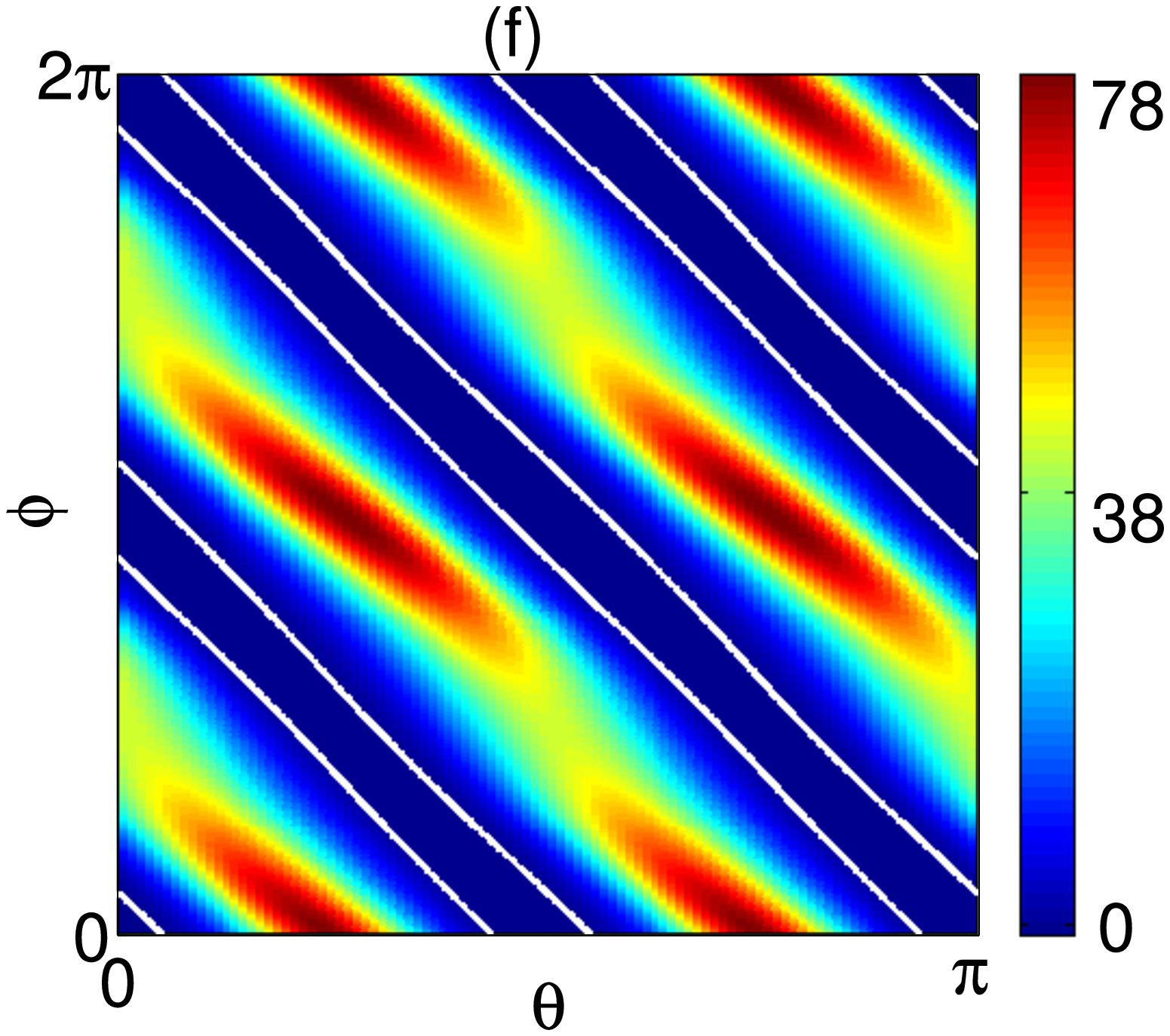} &
\hspace{-.25cm}\includegraphics[trim =7mm 0mm 0mm 0mm, clip, width=4.5cm]{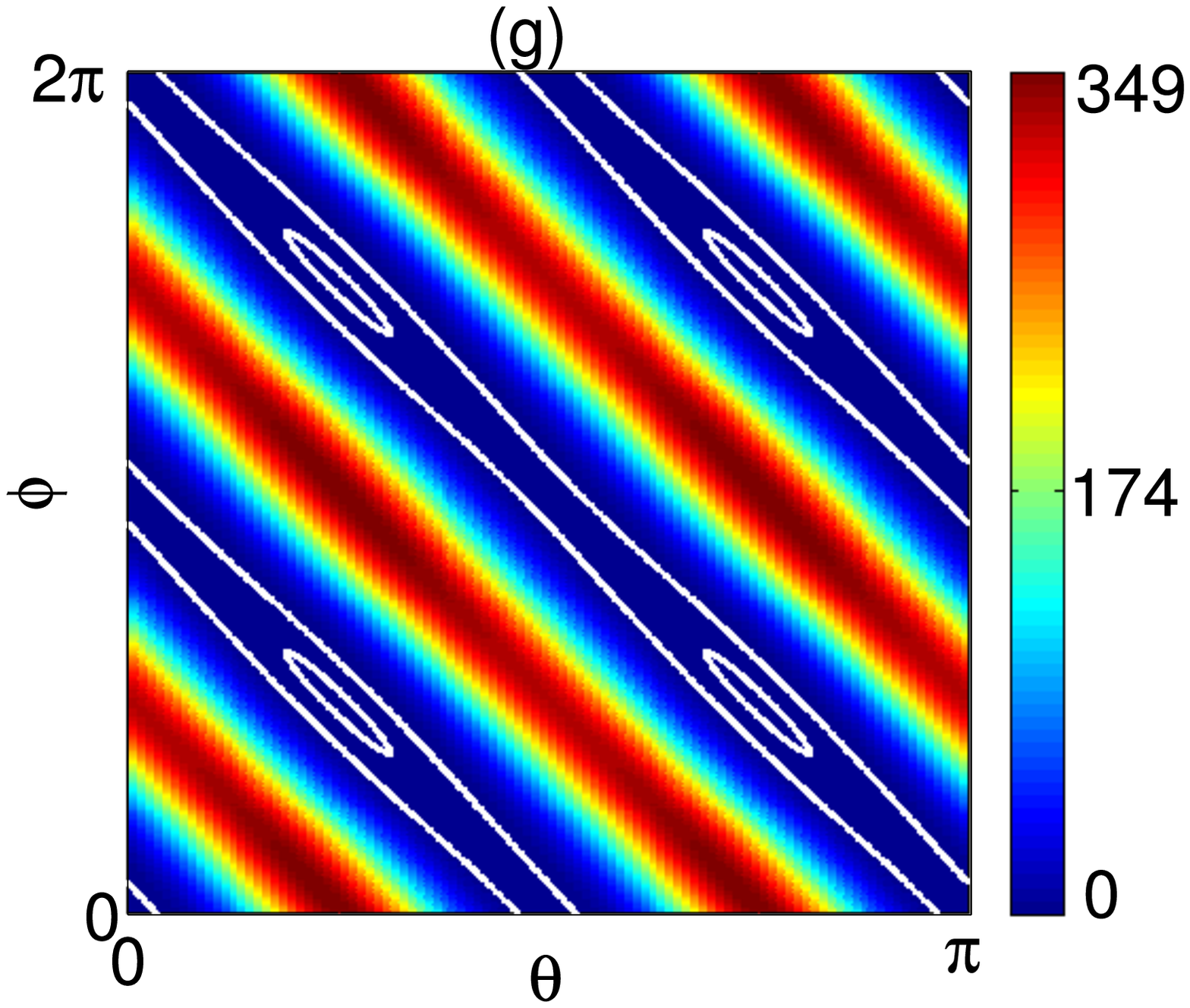} &
\hspace{-.25cm}\includegraphics[trim =7mm 0mm 0mm 0mm, clip, width=4.5cm]{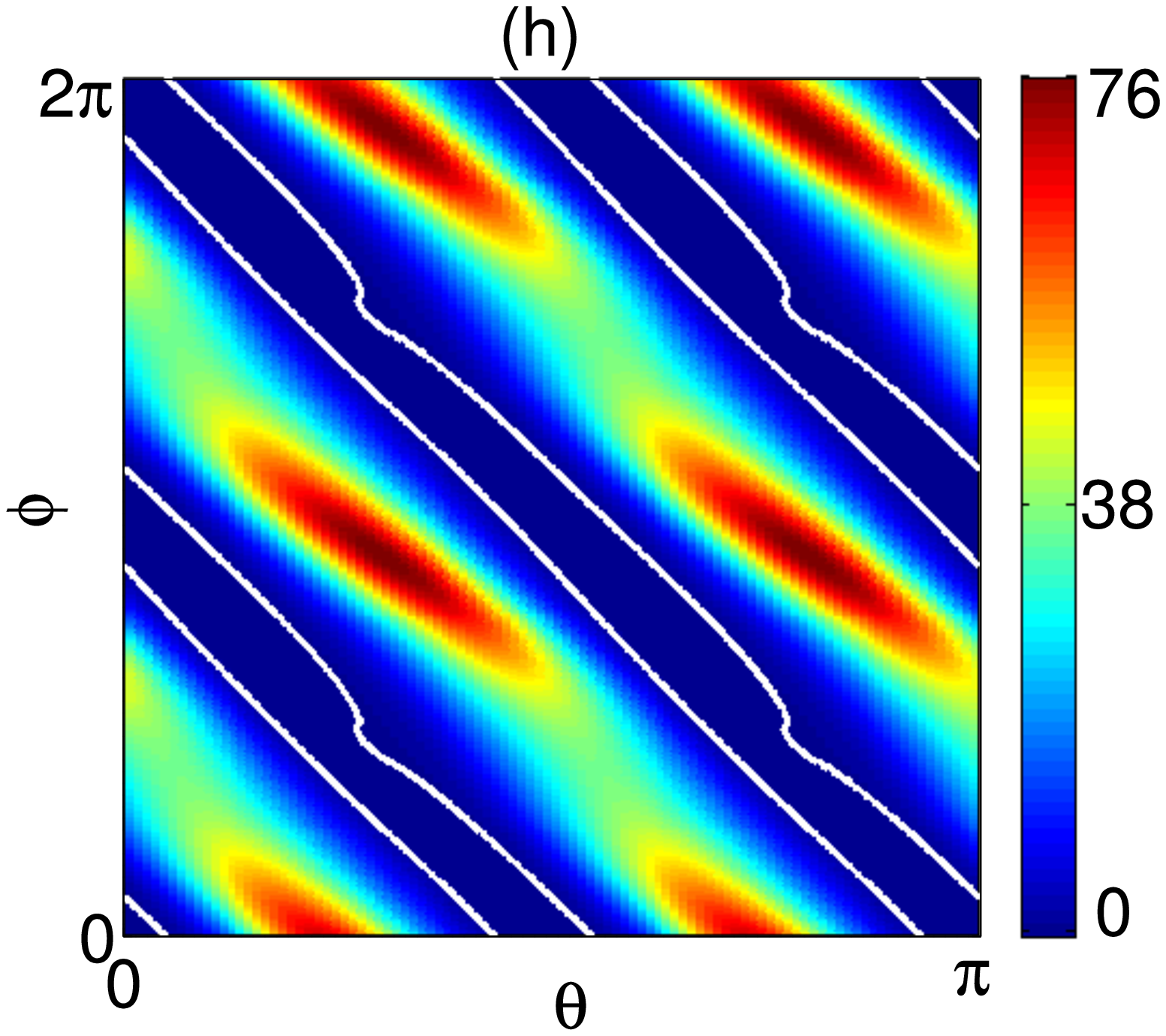}
\end{tabular}
\caption{(Color online) {\it Entanglement measured with inseparability and EPR-entanglement criteria   for different configurations of the PC}. (a) to (d) show the inseparability criteria taking $\sigma=1$ without PC and  for the cases $M_0=0.5$, $M_1=0$; $M_0=0$, $M_1=0.5$;  $M_0=0.5$,$M_1=0.5$, respectively. (e) to (h) show the EPR-entanglement criterion for the same cases.  All the cases are represented for a 5\%  below threshold. White lines demarcate the regions for which EPR entanglement is found according to both criteria.  \label{fig6}}
\end{figure*}

In order to check the validity of our approximations we now consider also
numerical results obtained from the full nonlinear and multimode model.
Fig.~\ref{fig5}  shows the variance for the fields for all the angles and different
configurations, comparing analytical and numerical results  obtained after simulation
of Eqs.~(\ref{Eq:lang}), leading to intracavity correlations. The largest squeezing for
an OPO in
absence of PC
is attained just at threshold between the critical modes while in the presence of the PC
this minimum value is changed. The agreement between numerical and analytical
calculations is very good. 

In general, we find that even considering an OPO and a PCOPO at a fixed distance
from the respective thresholds, the variance of $\Sigma_{\theta\varphi}$ is deeply
modified by the presence of the PC. While the achieved level of squeezing is not
affected by the  spatial modulation, the angles at which the maximum squeezing is reached are modified. On the other hand, we
mention that  the results presented in
Ref.~\cite{2011DeCastro} concerning the same device in the above threshold regime,
indicate strong effects such as the 
widening in the region of angles describing maximal squeezed states.

\subsection{Entanglement measures}

In this Section we analyze  the spatial entanglement between opposite modes for
different parameters in order  to verify if  the action of the refractive index
modulation disturbs entanglement measures and/or  creates separable states. It is
known that OPOs are useful devices giving spatially  entangled states generation below
threshold~\cite{1999GattiEPL,2003ZambriniPRA,spatialEPR}.  Let us   analyze in this regime how the modulation
introduced by the PC modifies quantum correlated states at opposite points in the far
field. First, we define the (momentum) quadrature operators $\hat{p}_i$.
These quadratures are defined similarly to
 $\hat{x}_i$, Eqs.~(\ref{eq:xi}), but with angles shifted as
$\theta\rightarrow \theta-\pi/2$ and $\varphi\rightarrow \varphi+\pi$.
Then, we introduce the following operators in order to establish if there is
Einstein-Podolsky-Rosen (EPR) entanglement~\cite{1935EinsteinPR}:
\begin{align*}
&\hat{u}=|\sigma|\hat{x}_1+\frac{1}{\sigma}\hat{x}_2\\
&\hat{v}=|\sigma|\hat{p}_1-\frac{1}{\sigma}\hat{p}_2
\end{align*}
where $\sigma$  is a real parameter.
The state inseparability criterion introduced in~\cite{2000DuanPRL} for continuous
variables systems, establishes that any separable quantum state characterized by a
density operator $\rho$ has a lower bound on the variances sum. With the notation
considered here, and introducing the parameter $\sigma$ in our definitions (\ref{x's}),
the inseparability criterion of Duan et al~\cite{2000DuanPRL} reads:
\begin{equation}\label{eq:duan}
 \Delta^2 \Sigma_{\theta\varphi}^\sigma+\Delta^2
 \Sigma_{\theta+\pi/2,\varphi+\pi}^\sigma\geq 2
 \left(\sigma^2+\frac{1}{\sigma}\right).
\end{equation}
In Fig.~\ref{fig6}a to d we show this sum of variances for all the relevant combinations of the 
quadrature angle $\theta$ and superposition angle $\varphi$ for four different
configurations. White lines identify the regions for which the bound~(\ref{eq:duan})
is violated, thus corresponding to entangled states. Our
analytical results show that this inseparability region is modified by a spatial
modulation, even when considering OPO and PCOPO operating at the same distance from the
threshold and that  the quadratures and superposition angular tolerance is slightly
 widened when the PC is introduced in the signal.

An alternative
criterion to characterize EPR entanglement in continuous variable systems was
proposed in Ref.~\cite{1988ReidPRL}. We introduce:
\begin{align*}
& \Delta^2 \Sigma_{\theta\varphi}^{\lambda}=\langle\left(\hat{x}_1+\lambda\hat{x}_2
\right)^2\rangle,
\end{align*}
minimized by
\begin{equation*}
\lambda=\frac{-\langle\left(\hat{x}_1\hat{x}_2\right)\rangle}
 {\Delta^2\hat{x}_2}.
\end{equation*}
According to EPR criterion given in~\cite{1988ReidPRL}, a state is EPR-entangled if
\begin{equation*}
 \Delta^2 \Sigma_{\theta\varphi}\Delta^2 \Sigma_{\theta+\pi/2,\varphi+\pi}\leq 1.
\end{equation*}
In Fig.~\ref{fig6}e to h  we show, for four different configurations,  the
analytical calculations for this quantity. White lines demarcate the region for which
the criterion is fulfilled and entanglement is found. Both measures show that the
region where entanglement can be found is slightly widened if the PC is introduced in
the signal. This effect is much relevant above threshold, as shown
in~\cite{2011DeCastro}.

\subsection{Twin beams correlations}

Finally, we present analytical results to characterize higher order correlations,
not related in a trivial way to entanglement~\cite{Giorgi}. We consider twin beams
correlations~\cite{1995Walls,1987HeidmanPRL} characterized by a negative value of the normal order variance ~$\langle
: \left( n (k)-n(-k)\right)^2 : \rangle $, where, according to the notation introduced
above, $n(k)$ is the fluctuation operator ($n=\hat N-\langle \hat N\rangle$) of the photon number
$\hat{N}(k)= \langle \hat{A}^{\dagger} (k,t)
\hat{A}(k,t) \rangle $. As we consider Gaussian states, all moments can be expressed as a
function of second order moments~\cite{1991Gardiner}, giving:
\begin{align}\nonumber
&\langle:( n(k)-  n(-k))^2 :\rangle= \langle a^\dagger(k)a(k)\rangle^2+\langle a^\dagger(-k)a(-k)\rangle^2\\ \nonumber
&+|\langle a(k) a(k)\rangle |^2+|\langle a(-k) a(-k)\rangle |^2-2|\langle a (k) a(-k)\rangle |^2 \\ 
&-2| \langle a (k) a^{\dagger}(-k)\rangle |^2.
\label{twin}
\end{align}

\begin{figure*}[t]
\begin{tabular}{ccc}
\includegraphics[width=5.75cm]{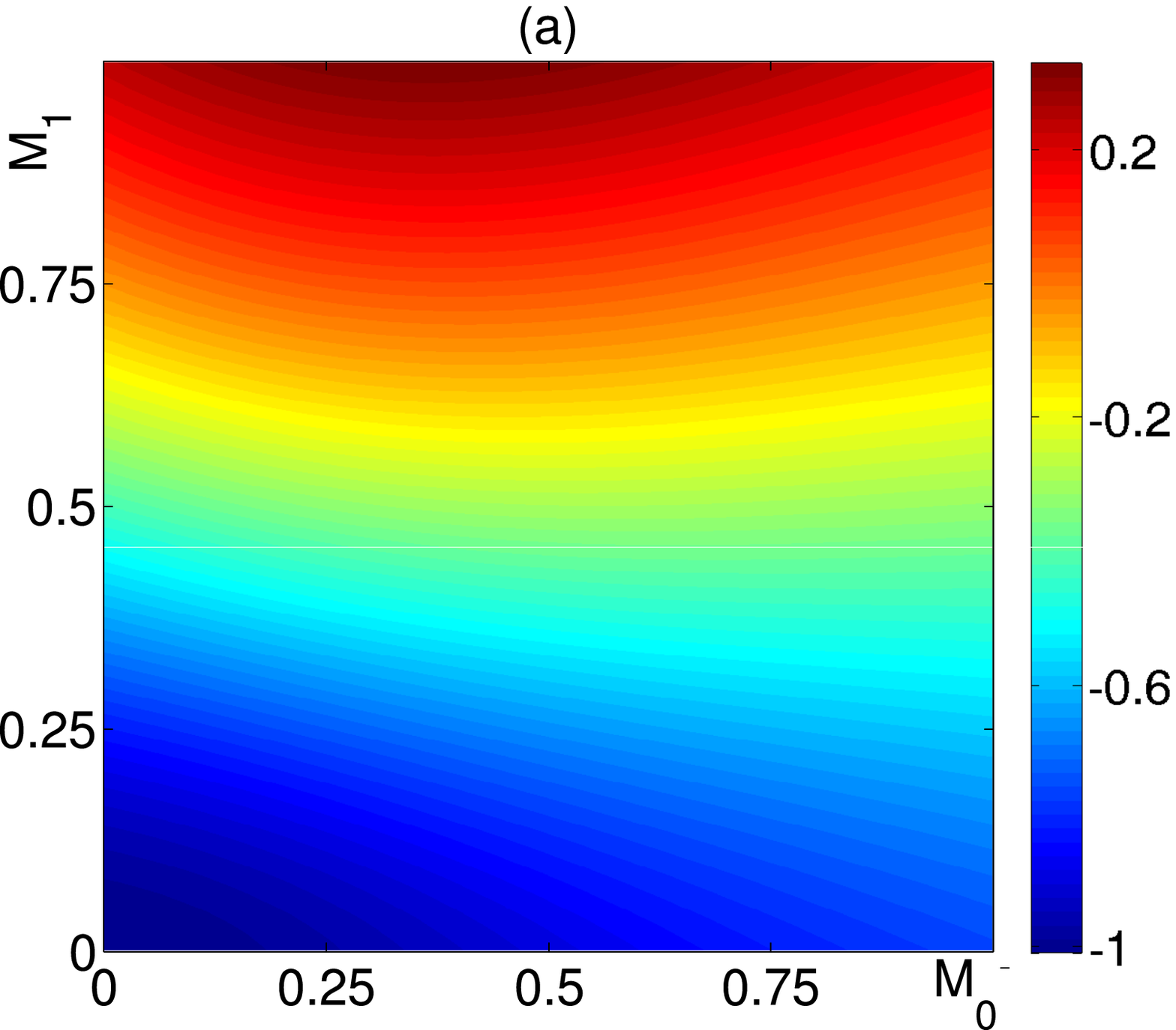} &
\includegraphics[width=5.75cm]{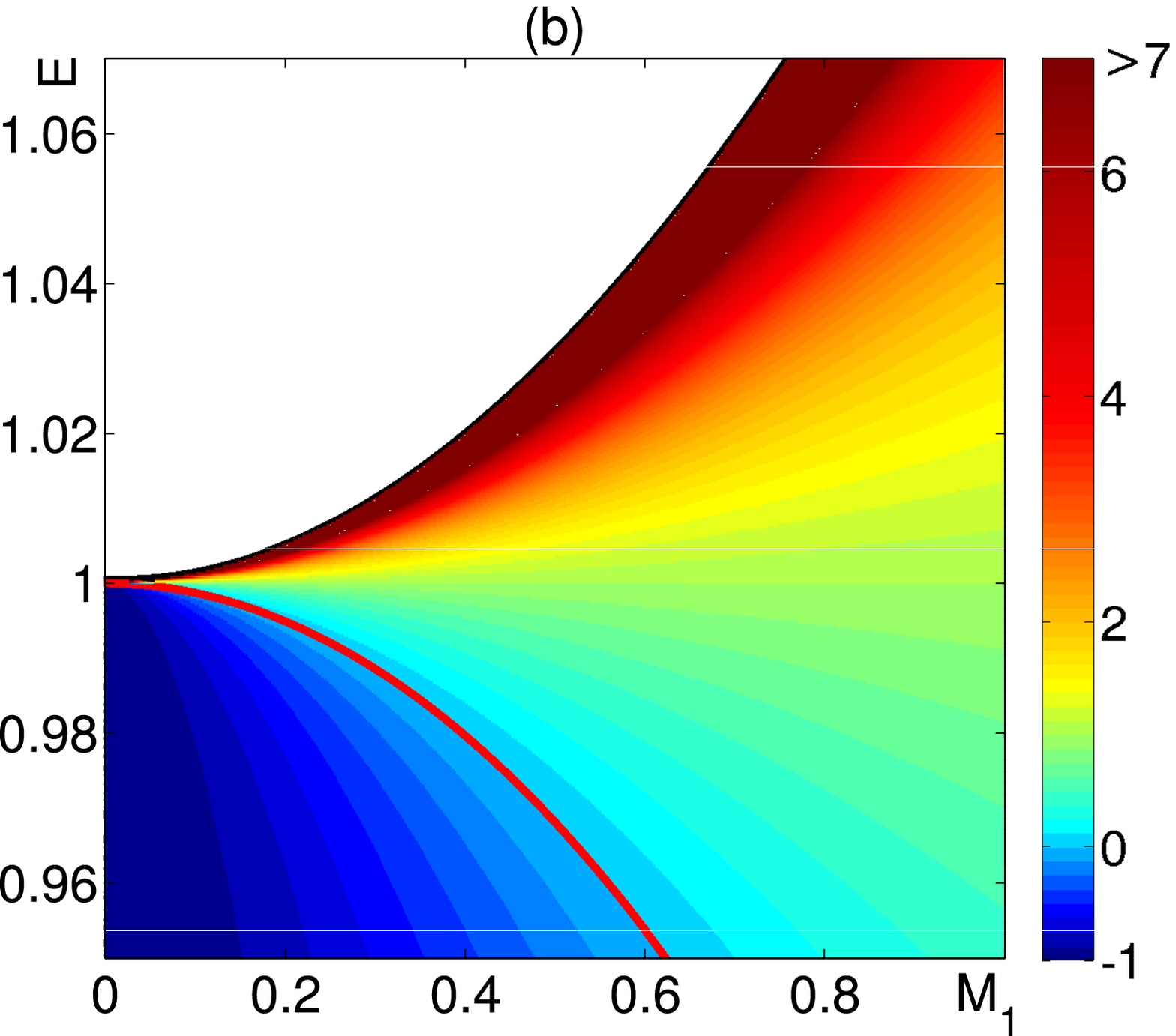} &
\includegraphics[width=6.cm]{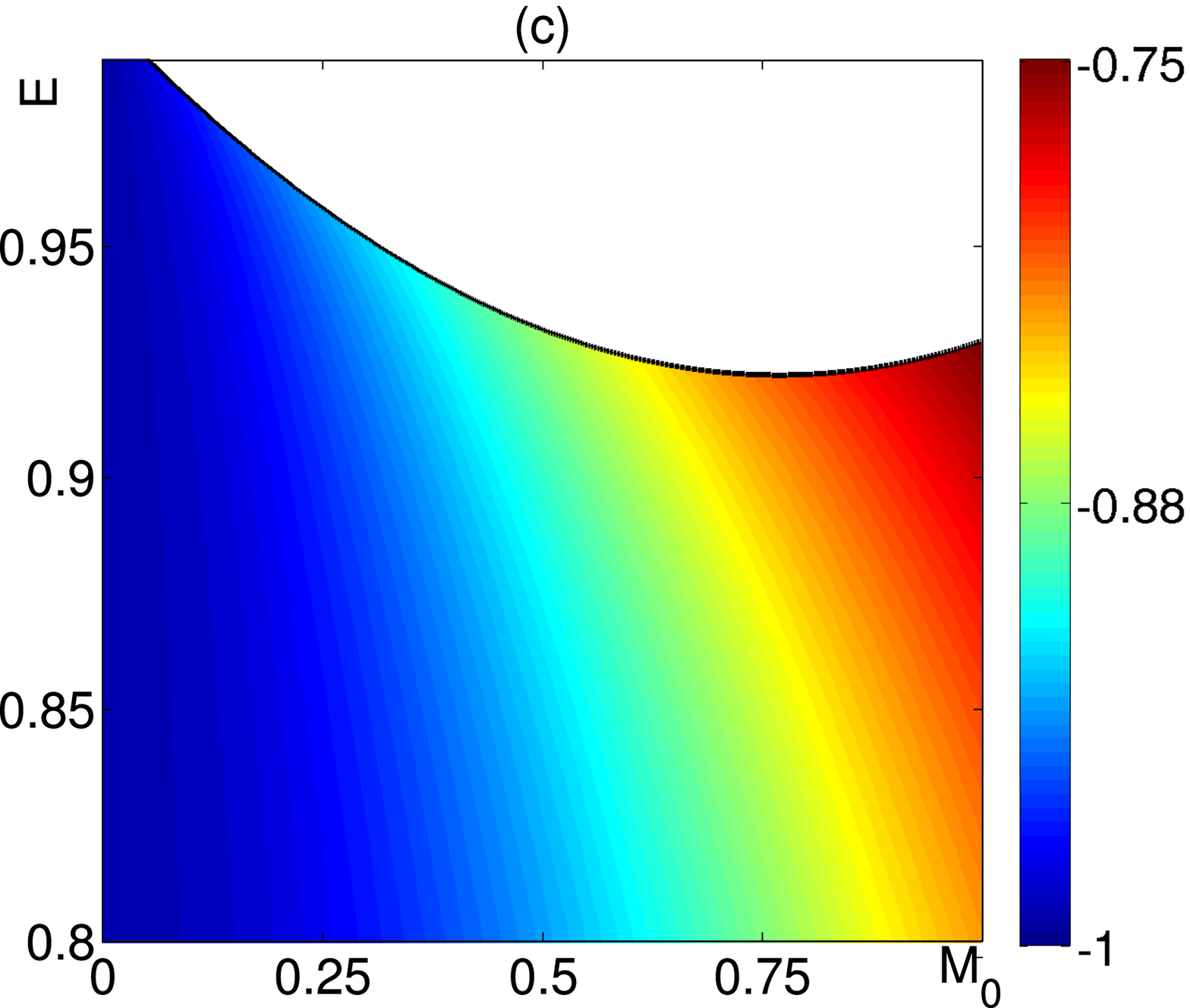}
\end{tabular}
\caption{(Color online) {\it Twin beam correlations obtained from the analytical expressions.} (a) Twin
beam correlations for different configurations of the PC and $E=0.92$. (b) Twin beam correlations below
threshold for different values of $M_1$ and $M_0=0$ and normalized with the shot noise. 
Black solid curve represents the value of the
threshold for every value of $M_1$. Red curve represents the change from negative to positive
(classical) correlation.  (c) same  for different values of $M_0$ and $M_1=0$. Black solid curve represents the value
of the threshold for every value of $M_0$. }\label{fig7}
\end{figure*}
The non classical feature of twin beams follows from the negativity of the variance
$\langle:( n(k)-n(-k))^2 :\rangle<0$ and from the corresponding singularity of the P
representation. Notice that terms $\langle a(k) a(k)\rangle$, in the second row of the
(\ref{twin})
vanish for an OPO in the absence of PC or when $M_1=0$. This is also evident from the
analytical
correlations given in the App.~\ref{sec:Correlationsw}.
Conversely, these terms are large in PCOPO whenever PC is affecting the signal,
i.e., for $M_1\neq0$. Moreover, the last term $\langle a (k) a^{\dagger}(-k)$
vanishes, both for $M_0=M_1=0$ and for
$M_0=0$ and $M_1\neq0$, but it can contribute to the variance negativity for
$M_0\neq0$.

In particular, for $M_0=M_1=0$, Eq.~(\ref{twin}) for the output fields yields the
simple expression
\cite{1997GattiPRA,Zambrini_twin}
\begin{equation}
\langle:( n(k_c)- n(-k_c))^2 :\rangle=-\frac{2E^2}{1-E^2}.
\end{equation}
with shot noise $ \langle  n(k_c)- n(-k_c)\rangle=2E^2/(1-E^2)$. Therefore the normalized variance is
always negative and constant below threshold. This it not the case for a PCOPO and in Fig.~\ref{fig7}a we
show twin beams correlations normalized by the shot noise for  a fixed value of the pump and different PC
modulations.  The OPO is recovered at the origin, while the twin beams correlation is modified by the PC, 
becoming more classical for higher values of $M_1$.  Let us consider now the case of modulation  only in
the signal detuning, that is $M_0=0$ and $M_1\neq0$. We find for the twin beams correlations
\begin{equation}\label{twin-m1}
\langle:( n(k_c)-n(-k_c))^2 :\rangle=\frac{8E^2(4E^2+M_1^2-4)}{(4-4E^2+M_1^2)^2}.
\end{equation}
with shot noise $ \langle n(k_c)- n(-k_c)\rangle=8E^2/(4-4E^2+M_1^2)$. The normalized expression of the
twin beams diverges for $4-4E^2+M_1^2=0$, which in turn is the expression for the threshold,
Eq.~(\ref{Eq:thresh}) (black solid curve in Fig.~\ref{fig7}b).
 We find that, even after normalization with the shot noise, the strength of these correlations is
dependent on all parameters, while in the OPO it was constant for any pump E. Moreover, in  
Fig.~\ref{fig7}b it is clear that the correlations are degraded and can become classical below threshold
(above red line).
Finally, in  Fig.~\ref{fig7}c we present   normalized  twin
beams correlations when only the pump detuning is modulated. In this case correlations remains quantum
for all pump strengths below threshold, even if there is a small reduction of this quantum effect. 

An interpretation of these results can be given considering the modulation effects from the microscopic
point of view. A modulation in the signal  (pump) detuning has the main effect to couple different spatial
modes as also clear from Eq.(\ref{Eq:enFourier}). In the particular case of a PC modulation with the
periodicity (\ref{2kc}) there are actually creation and destruction processes between photons at the
critical tilted modes $\pm k_c$. Even if the nonlinearity gives rise to simultaneous creation
(annihilation) of photons pairs in the signal,  spatial modulation of the signal detuning  leads to
photons hopping between these opposite modes. In other words there is a process of creation of one photon
(say $a_1(-k_c)$) and destruction of one in the opposite mode ($a_1^\dagger(k_c)$), as can be also seen
from inspection of the Hamiltonian (\ref{Eq:Hamiltonian}).  Therefore, twin beams correlations, due to
photon pairs emission, are present for small hopping rates, but when the detuning  modulation of the
signal is increased, the variance (\ref{twin-m1}) becomes classical as the twin beams are depleted
uncoherently. On the other hand, for $M_1=0$ and
modulating the detuning  of the pump field, the hopping between different pump harmonics is still
detrimental, as it triggers different secondary processes besides the twin photons pairs
generation, but has a reduced effect.

\section{Conclusions}

We have shown the effect of an intracavity PC in a typical device displaying quantum
light spatially correlated in continuous variables, as it is the degenerate OPO type
I. Due to the translational symmetry breaking in the transverse profile there are
several effects also evident in the noisy precursors locking (Fig.~\ref{fig2}). The PC
modulation has a deep influence in the instability process and parametric threshold
are both raised and lowered depending on the configuration (Fig.~\ref{fig5}). As a
consequence a  modification in twin beams correlations, squeezing, separability, and
entanglement were expected. In order to analytically evaluate these quantum
correlations we have introduced two main approximations valid  below threshold,
linearizing around the steady state where the signal field vanishes and
restricting our analysis to few relevant harmonics (see  Fig.~\ref{fig2}). We have
considered a PC modulation (\ref{2kc}) and five modes,  $k=0,\pm k_{\mathrm{pc}}$ for the pump and $k=\pm k_{\mathrm{pc}}$ for the
signal. Under these approximations we obtained good agreement between our results and
the numerical solution of the full multimode nonlinear model,  Eqs.~(\ref{Eq:lang}).

In our prototype model the PC leads to a sinusoidal variation on the refractive index
and, therefore, of the detunings that can affect pump, signal or both fields,
depending on the values of the PC modulations $M_0$ and $M_1$. Different
configurations are described by tuning these parameters. When the PC modulates the
signal field, i.e. $M_1\neq0$, the instability threshold rises in agreement with
previous works concerning single resonant cavities and predicting pattern
inhibition~\cite{2004GomilaPRL,2005GomilaPRE}. We demonstrate that in presence of a
parametric process the scenario is more complex and actually the threshold can also be
reduced, when $M_0\neq0$, and the instability favored.

Non-classical phenomena such as squeezing and entanglement are very sensitive to the
proximity of the instability point. Therefore the PC has deep consequences and can
improve correlations at a given pump with respect to the OPO, at least when threshold
is lowered (pump modulation). Apart from this effect, all  correlations have been
compared in OPO and PCOPO at a  fixed distance from threshold.  Then we have found
that squeezing (Fig.~\ref{fig4}), separability, and EPR-entanglement (Fig.~\ref{fig6})
are preserved both in the reached values and in the width of the quadrature and
superposition angles regions where these phenomena appear.  Finally, we have
analytically calculated twin beams correlations
(Fig.\ref{fig7}) when varying the PC modulations $M_0$ and $M_1$ showing that deeper effects are
actually present in these intensity correlations, even at a fixed distance from the threshold, and that in general secondary
processes degrade correlations.
 Besides, numerical simulations above threshold presented in a previous
work~\cite{2011DeCastro} have revealed a significant enhancement of squeezing and
entanglement. Overall, the PC allows to obtain the same quantum effects at a lower
energy and one can enhance/avoid quantum properties of light, just by changing pump
and/or signal spatial modulation in the cavity.

\appendix
\section{Linear dynamics of PCOPO with any $k_p$}
\label{Sec:BigL}

If we Fourier transform in the temporal variable Eqs.~(\ref{Eq:enFourier}) and neglect all the terms with $|k|>k_{\mathrm{p}}$ in the signal and pump, we obtain
\begin{equation}
\label{Eq:LEqs}
\bar{L}_6\vec{a}_{1}=\sqrt{\frac{2}{\gamma}}\vec{a}_{1}^{\mathrm{in}},
\end{equation}
where $6$ modes are coupled between them, namely
\begin{eqnarray*}
\vec{a}_{1} &=&  \Big(a_{1}(k),a_{1}(k+k_{p}),  a_{1}(k-k_{p}),\\
& & \left. a_{1}^{\dagger}(-k),a_{1}^{\dagger}(-k-k_{p}),a_{1}^{\dagger}(-k+k_{p})\right)^{\top}.
\end{eqnarray*}
We use a  compact notation  and denote $ a_{1}(k,\omega)$ simply as $ a_{1}(k)$, $ a_{1}^{\dagger}(k,-\omega) $ as $a_{1}^{\dagger}(k)$, and similarly for
$\vec{a}_{1}^{\mathrm{in}}$.
Let us call $ \eta(nk_p)=-i\omega+\left(1+i\Delta_{1}+i2(k+n k_p)^{2}\right)$, $ \eta'(nk_p)=-i\omega+\left(1-i\Delta_{1}-i2(k+n k_p)^{2}\right)$,  $ S=\tilde{S}_{0}^{s}(0)$, and
$ \bar{\kappa}=(-M_{0}/2)/(1+ik_{p}^{2}).$ Then we can write the matrix $\bar{L_6}$ as
\begin{eqnarray*}
\left(\begin{array}{cccccc}
\eta(0) & -\frac{M_{1}}{2} & \frac{M_{1}}{2} & -S & \bar{\kappa} S & -\kappa S \\
\frac{M_{1}}{2} & \eta(k_{p}) & 0 & -\bar{\kappa} S & -S & 0\\
-\frac{M_{1}}{2} & 0 & \eta(-k_{p}) & \bar{\kappa} S & 0 & -S \\
-S^{*} & -\bar{\kappa}^{*}S^{*} & \bar{\kappa}^{*}S^{*} & \eta'(0) & \frac{M_{1}}{2} & -\frac{M_{1}}{2}\\
\bar{\kappa}^{*}S^{*} & -S^{*} & 0 & -\frac{M_{1}}{2} & \eta'(k_{p}) & 0\\
-\bar{\kappa}^{*}S^{*} & 0 & -S^{*} & \frac{M_{1}}{2} & 0 & \eta'(-k_{p})\end{array}\right).\end{eqnarray*}
This matrix allows the dynamical description of the fluctuations for all modes such that
$|k|\le k_{p}$, with continuous index $k$.

When considering the modulation $k_p=k_c/2$ with $k_c$ critical wavenumber, few intense modes are relevant and a reduced description can be obtained, leading to
the $4\times 4$ matrix given in Eq.~(\ref{Eq:matL}).

\section{Solution of the input-output equation}
\label{sec:InverseL}

The output fields dynamics is governed by the Eqs. (\ref{Eq:In-out}).
The inverse of $L$ is:
 \begin{eqnarray*}
\frac{1}{D(\omega)}\left(\begin{array}{cccc} U(\omega) & V(\omega) & W(\omega) &
Z(\omega)\\ -V(\omega) & U(\omega) & -Z(\omega) & W(\omega)\\ W'(\omega) & Z'(\omega)
& U'(\omega) & V'(\omega)\\ -Z'(\omega) & W'(\omega) & -V'(\omega) &
U'(\omega)\end{array}\right),\end{eqnarray*} where $D(\omega)$ is the determinant %
\begin{eqnarray*} D(\omega)  =
\frac{1}{2}c_1c_2+\frac{1}{4}c_2^2+|S|^{4}|1+\kappa^{2}|^2,\end{eqnarray*} with
$c_1=2|S|^2(1+\kappa^2)$, $c_2=-\frac{1}{2}(M_1^2-4(i+\omega)^2)$. Finally, the matrix
components are

\begin{eqnarray*}
 U(\omega) &=& -\frac{1}{2}(c_1+c_2)^2-(1+i\omega)(c_1+c_2)+M_1c_3,\\
 U'(\omega) &=& -\frac{1}{2}(c_1+c_2)^2-(1+i\omega)(c_1+c_2)-M_1c_3,\\
 V(\omega) &=& c_4+\frac{M_1}{2}(c_1+c_2),\,\,\,
 V'(\omega) = c_4-\frac{M_1}{2}(c_1+c_2), \\
 W(\omega) &=& -S(c_1+c_2), \,\,\,
 W'(\omega) = -S^*(c_1^*+c_2), \\
Z(\omega) &=& -S(c_1 \kappa^*+c_2 \kappa), \,\,\,
Z'(\omega) = S^*(c_1^*\kappa+c_2\kappa^*),
\end{eqnarray*}
where  $c_3=|S|^2(\kappa-\kappa^*)$  and $c_4=2c_3(1-i\omega)$.

\section{Second order moments in frequency and time domains}
\label{sec:Correlationsw}

We can use the expression of the output variables in terms of the input ones given in
Eq.~(\ref{Eq:In-out}), to calculate different second order correlations in the frequency domain. For example, we can consider the correlation at different $\omega$ and the same $k$ between $a^{\mathrm{out},\dagger}$ and $a^{\mathrm{out}}$. To solve the expression obtained for these correlations in terms of the input variables, we should consider that
$
\left\langle a^{\mathrm{in}}(k,\omega)\, a^{\mathrm{in},\dagger}(k',\omega')\right\rangle =\delta(k-k')\delta(\omega-\omega'),$
while any other combination vanishes [this property is obtained  Fourier transforming the expression
$\left\langle a^{\mathrm{in}}(k,t)\, a^{\mathrm{in},\dagger}(k',t')\right\rangle =\delta(k-k')\delta(t-t')$]. For example, we have:
\begin{align*}
&  \left\langle a^{\mathrm{out},\dagger}(k_c,\omega)\,a^{\mathrm{out}}(k_c,\omega')\right\rangle =\\
&  4\frac{W'(\omega)W(-\omega')-Z'(\omega)Z(-\omega')}{D(\omega)D(-\omega')}\delta(0)\delta(\omega-\omega').
\end{align*}
We obtain $\delta(\omega-\omega')$ from the Fourier transform definition.
So $\omega'=\omega$ due to this delta function. Notice that the matrix
offers the term $a^{\mathrm{out},\dagger}(k_c,-\omega')$, so to
obtain $a^{\mathrm{out},\dagger}(k_c,\omega')$ it is necessary
to change the sign of the terms in the matrix. Then, for the spectral intensity we obtain:
\begin{align}
\label{eq:espectI}
 \left\langle a^{\mathrm{out},\dagger}(k_c,\omega)\,a^{\mathrm{out}}(k_c,\omega)\right\rangle =\nonumber\\
4\frac{W'(-\omega)W(\omega)-Z'(-\omega)Z(\omega)}{|D(\omega)|^{2}}.
\end{align}
To obtain the intensity in the time domain we should Fourier transform the previous expression:
\begin{align*}
& \left\langle a^{\mathrm{out},\dagger}(k_c,t)\,a^{\mathrm{out}}(k_c,t)\right\rangle   = \\
  & \frac{1}{2\pi}\int d\omega d\omega'e^{i(\omega'-\omega)t}\left\langle a^{\mathrm{out},\dagger}(k_c,\omega),a^{\mathrm{out}}(k_c,\omega')\right\rangle\\
  & =  \frac{2}{\pi}\int d\omega\frac{W'(-\omega)W(\omega')+Z'(-\omega)Z(\omega')}{|D(\omega)|^{2}}.
\end{align*}
To solve this integral we have to consider that the denominator shows eight poles of the form:
\begin{equation*}
\omega=\pm i \pm \sqrt{-|S|^{2}+i|S|^{2}(\kappa-\kappa^{*})-|S|^{2}|\kappa|^{2}+M_{1}^{2}},
\end{equation*}
and we note that there are four in the upper part of the complex plane and four in the lower part. Performing this integral using conventional methods we obtain~(\ref{Eq:Intensity}). Any other second order correlation can be obtained using Eq.~(\ref{Eq:In-out}) as described above. For example, all the non vanishing terms in the expression of the variance $\Delta^2 \Sigma_{\theta\varphi}$    are:
\begin{align*}
&\langle\hat{x}_1^2\rangle= 2\Re \left\langle a\left(k_c\right) a\left(k_c\right)\right\rangle e^{2i\theta}\\
 & +2\left\langle a^{\dagger}\left(k_c\right) a\left(k_c\right)\right\rangle+\left[a^{\dagger}\left(k_c\right),a^{\dagger}\left(k_c\right)\right],
\end{align*}
\begin{align*}
&\langle\hat{x}_2^2\rangle= 2\Re \left\langle a\left(-k_c\right) a\left(-k_c\right)\right\rangle e^{2i(\theta+\varphi)}\\
 & +2\left\langle a^{\dagger}\left(-k_c\right) a\left(-k_c\right)\right\rangle+\left[a^{\dagger}\left(-k_c\right),a^{\dagger}\left(-k_c\right)\right],
\end{align*}
\begin{align*}
&\langle\hat{x}_1\hat{x}_2\rangle=2\Re \left\langle a\left(k_c\right) a\left(-k_c\right)\right\rangle e^{i(2\theta+\varphi)}\\
& + 2\Re \left\langle a^{\dagger}\left(k_c\right) a\left(-k_c\right)\right\rangle e^{i\varphi}.
\end{align*}
Finally, the Fourier transforms of all the relevant correlations used in the definition of the variance are:
\begin{align*}
 \left\langle a^{\mathrm{out},\dagger}(-k_c,t)\, a^{\mathrm{out}}(k_c,t)\right\rangle & = 4 c_3 c_5/\sigma,
\end{align*}
\begin{align*}
 \left\langle a^{\mathrm{out}}(k_c,t)\, a^{\mathrm{out}}(-k_c,t)\right\rangle  & = \left(2 S (-2c_1)c_6^*+c_5 c_6)\right)/\sigma,
\end{align*}
\begin{align*}
\left\langle a^{\mathrm{out}}(k_c,t)\, a^{\mathrm{out}}(k_c,t)\right\rangle & = -\left(2 S (2 c_1 c_7^*- c_5 c_7)\right)/\sigma,
\end{align*}
\begin{align*}
\left\langle a^{\mathrm{out},\dagger}(k_c,t)\, a^{\mathrm{out},\dagger}(-k_c,t)\right\rangle & = \left(2 S^{*}(-2 c_1^*c_6+c_5 c_6^*\right)/\sigma,
\end{align*}
\begin{align*}
\left\langle a^{\mathrm{out},\dagger}(k_c,t)\, a^{\mathrm{out},\dagger}(k_c,t)\right\rangle & = -\left(2 S^{*}(2 c_1^*c_7-c_5 c_7^*)\right)/\sigma,
\end{align*}
where $\sigma$ and $c_1$ were defined above, $c_5=(4+M_{1}^{2})$, $c_6=2+\kappa M_{1}$, and $c_7=2 \kappa -M_{1}$.

\end{document}